\documentclass[float,12pt]{article}

\usepackage{epsf}
\usepackage{graphicx,epsfig}
\usepackage{amsfonts}
\usepackage{amssymb}
\usepackage{cite}

\makeatletter
\renewcommand\section{\@startsection {section}{1}{\z@}%
                                 {-3.5ex \@plus -1ex \@minus -.2ex}%nn
                                   {2.3ex \@plus.2ex}%
                                   {\normalfont\large\bfseries}}
\renewcommand\subsection{\@startsection{subsection}{2}{\z@}%
                                   {-3.25ex\@plus -1ex \@minus -.2ex}%
                                     {1.5ex \@plus .2ex}%
                                     {\normalfont\bfseries}}
\makeatother

\newcommand{\Letter}{
\setlength{\textwidth}{16.5cm} 
   \setlength{\textheight}{22.6cm} 
    \hoffset=-0.75in 
\voffset=-2.1cm }
    
\Letter

\makeatother

\newcommand{\eps}{\epsilon}   
      \renewcommand{\th}{\theta}

\newcommand{\Sig}{\Sigma}

    \newcommand{\cR}{{\cal R}}
\newcommand{\cS}{{\cal S}}

\newcommand{\vac}{|0\rangle}

\def\preal{{\rm Re\,}}
\def\pim{{\rm Im\,}}

\def\yzero{\smash{\hbox{$y\kern-4pt\raise1pt\hbox{${}^\circ$}$}}}
\def\p{\partial}
\def\a{\alpha}

\def\g{\gamma}
\def\d{\delta}
\def\beq{\begin{equation}}
\def\eeq{\end{equation}}
\def\beqa{\begin{eqnarray}}
\def\eeqa{\end{eqnarray}}
\def\Om{\Omega}
\def\om{\omega}
\def\th{\theta}

\def\-{\hphantom{-}}

\def\s2{\frac{1}{\sqrt2}}

\def\oh{\frac{1}{2}}

\def\Tr{{\rm Tr \,}}

\def\vac{|0 \rangle}

\def\cn{{\cal N}}
\def\cam{{\cal M}}

\def\Dsl{\,\raise.15ex\hbox{/}\mkern-13.5mu D} %this one can be subscripted
\def\IZ{Z\kern-.4em  Z}
\def\id{{\rm 1}}

\def\ta{\tilde{A}}
\def\tb{\tilde{B}}

\def\eps{\epsilon}
\def\k{\kappa}

\def\car{{\cal R}}
\def\l{\lambda}
\def\raw{\rightarrow}
\def\Raw{\Rightarrow}
\def\inte{{\bf Z}}
\def\cpx{{\bf C}}

\def\T{{\bf T}}

\newcommand{\be}{\begin{equation}}
\newcommand{\ee}{\end{equation}}
\newcommand{\bea}{\begin{eqnarray}}
\newcommand{\eea}{\end{eqnarray}}

\begin{document}
\thispagestyle{empty}
\begin{flushright}
\parbox[t]{1.2in}{MAD-TH-04-6}
\end{flushright}

\vspace*{.5in}

\begin{center}
{\large \bf Model Building 
and
Phenomenology of 
Flux-Induced \\
\vspace{3mm}
Supersymmetry Breaking on D3-branes}

\vspace*{0.5in} 
{Fernando Marchesano$^{1}$, Gary Shiu$^1$, and Lian-Tao Wang$^{1,2}$}
\\[.3in]
{\em  $^1$Department of Physics,
  University of Wisconsin,
Madison, WI 53706, USA}\\
{\em $^2$Jefferson Laboratory of Physics, Harvard University, Cambridge, MA 02138, USA} \\ [0.5in]
\end{center}

\begin{center}
{\bf
Abstract}
\end{center}

We study supersymmetry breaking effects induced on D3-branes at singularities by the presence of NSNS and RR 3-form fluxes. First, we discuss some local constructions of chiral models from D3-branes at singularities, as well as their global embedding in flux compactifications. The low energy spectrum of these constructions contains features of the supersymmetric Standard Model. In these models, {\it both} the soft SUSY parameters and the $\mu$-term are generated by turning on the 3-form NSNS and RR fluxes. We then explore some model-independent phenomenological features as, e.g., the fine-tuning problem of electroweak symmetry breaking in flux compactifications. We also comment on other phenomenological features of this scenario.

\vfill

\hrulefill\hspace*{4in}

{\footnotesize
Email addresses: marchesa@physics.wisc.edu, shiu@physics.wisc.edu, liantaow@schwinger.harvard.edu.}

\newpage

\tableofcontents

\vspace*{1.5cm}

\section{Introduction}

Despite the tremendous insight that string theory has provided us in understanding field theory phenomena, most of the progress so far is based to a large extent on the assumption of exact supersymmetry. Yet, to recover the observed low energy physics, supersymmetry must somehow be broken. 
An important open problem in string theory is therefore {\it how} to break supersymmetry while maintaining its beneficial features. The precise way in which supersymmetry is broken is also of great phenomenological importance because current and upcoming experiments at the Tevatron or the LHC will soon begin to probe the existence of superpartners and possibly the pattern of their soft masses.
Over the past few years, it has become clear that the phenomenological possibilities of string theory increase immensely with the inclusion of branes and fluxes. For example, in the brane world scenario, supersymmetry can be broken either in the D-brane sector or in the bulk. In the latter case, the D-brane sector preserves ${\cal N}=1$ supersymmetry, whereas supersymmetry breaking effects in the closed string bulk will be transmuted to the observable open string Standard Model sector via gravitational interactions. From an effective field theory point of view, the effects of supersymmetry breaking in the bulk felt by the Standard Model can be described by a set of soft parameters in the resulting $D=4$, ${\cal N}=1$ supergravity Lagrangian.

In this paper, we will focus on soft SUSY-breaking induced by type IIB 3-form NSNS and RR fluxes \cite{flux1,drs,flux2,gkp}, which is an example of bulk supersymmetry breaking.  An appealing feature of this scenario is that one can perform rather explicit string theoretical computations of the soft parameters \cite{ciu,ggjl,grana,lrs,ciu2}. More precisely, the 3-form fluxes couple to the Dirac-Born-Infeld (DBI) action describing the degrees of freedom on the D-branes and hence one can deduce the soft parameters from the relevant operators of the DBI action.
In addition to inducing soft SUSY-breaking terms, flux compactifications have also shown to play an important role in stabilizing moduli \cite{drs,gkp,kst}, as well as for constructing metastable dS vacua \cite{kklt} and inflationary models \cite{kklmmt,inflation,velocity} 
from string theory. 

As a first step towards studying the phenomenology of flux-induced SUSY breaking scenario, we need to construct chiral models in flux compactifications whose low energy spectrum contains semi-realistic features. Our framework to accomplish this task will be Type IIB orientifold compactifications involving D-branes. There are two known ways in which non-Abelian gauge groups and chiral fermions charged under them can arise in this context, either by considering D-branes localized at singularities \cite{firstquiver} or magnetized D-branes \cite{bachas,bgkl,aads,magnus}. The latter are related to intersecting D-branes constructions by T-duality \cite{bdl}.\footnote{Magnetized/Intersecting D-branes and D-branes an singularities are not string theory constructions which exclude each other. Indeed, as shown in \cite{afiru} both can be combined to give new chiral models.} In this work, we will restrict ourselves to D3-branes at singularities, \footnote{For recent work on constructing MSSM flux vacua from magnetized D-branes, see \cite{ms}.} since the soft SUSY-breaking effects are much easier to work out. In fact, the soft SUSY-breaking terms on D3-branes at singularities in a general background with fluxes have recently been derived in \cite{ciu,ggjl}, following the approach in \cite{grana}. The reason for the simplicity of this setup is that the Standard Model is localized on a single stack of D3-branes, filling four non-compact dimensions and being pointlike in an internal manifold $\cam_6$. Hence, one can deduce the flux-induced soft parameters by performing a {\it local} expansion of the supergravity background around the location of the D3-branes.\footnote{For recent developments on SUSY breaking soft-terms in systems of D3/D7-branes see \cite{lrs,ciu2}} We will present some global constructions of chiral models as explicit realizations of this scenario.

Another appealing feature of flux-induced SUSY-breaking is that all the soft parameters and the $\mu$-term are induced by different components of the 3-form flux. Hence, we have in principle all the necessary ingredients to explore issues in electroweak symmetry breaking (EWSB). However, it turns out that in the semi-realistic D3-brane models constructed in this context so far \cite{cu,ciu,throat}, the would-be $\mu$-term is absent. This is because the necessary component of the fluxes which induce the $\mu$-term is projected out by the symmetry of the background (more specifically, the orbifold symmetry). Thus, the purpose of this work is to construct and study some concrete examples of flux compactifications whose low energy spectrum contains features of the Standard Model, and which allow for fluxes that induce non-zero soft SUSY parameters and a non-vanishing $\mu$-term. We also demand that the models are free of extra chiral exotics charged under the SM gauge group and other sources of supersymmetry breaking, so that the resulting phenomenological features are due entirely to the fluxes.
Note that unlike in the magnetized D-brane setup, in which a three-generation MSSM model has been constructed \cite{ms}, it is rather difficult to construct a three-generation MSSM flux vacua from only D3-branes at singularities and which also allows for a $\mu$-term. In fact, the simplest model we found in this setup has only two chiral families. Nonetheless, as long as we explore issues which are insensitive to the number of families, we expect the results to be generic and applicable when more realistic models can be constructed in the future.

This paper is organized as follows. In Section \ref{ModelBuilding}, we construct some chiral models of flux compactifications from D3-branes at singularities.  Our strategy is to first construct a local singularity where a semi-realistic model can be localized, and which admits fluxes leading to {\it both} the soft SUSY-breaking terms and the $\mu$-term. We then embed such a local singularity into a fully-fledged compactification. In Section \ref{Phenomenology}, we explore some phenomenological features of flux-induced supersymmetry breaking and discuss the fine-tuning problem of EWSB in this scenario. We conclude with some outlook in Section \ref{Conclusion}. Some details about fluxes and NSNS tadpoles, as well as the quantization conditions for the 3-form fluxes are relegated to the appendices. 

\section{Model Building}\label{ModelBuilding}

\begin{figure}[ht]
\centering
\epsfxsize=5.5in
\epsffile{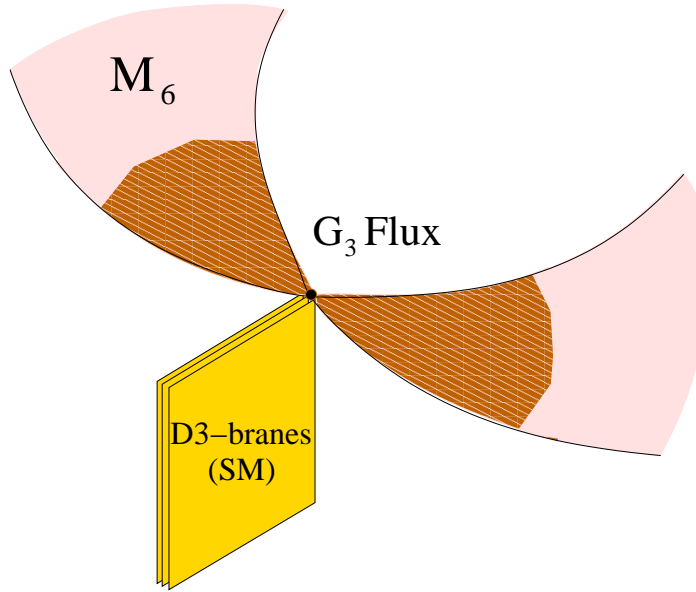}
\caption{\small{Flux-induced Supersymmetry Breaking scenario, for the particular case of D3-branes at singularities. The D3-branes sit at a singularity of the internal space $\cam_6$, in order to achieve a chiral gauge theory. The geometrical background preserves $\cn=1$ supersymmetry but the 3-form flux $G_3$ surrounding the singularity breaks it, inducing soft terms in the gauge theory.}}
\label{fluxD3}
\end{figure}

In this section, we construct some simple models where the general flux-induced soft SUSY-breaking scenario discussed in \cite{ciu} (see also \cite{ggjl}) can be realized. In order to apply the formalism developed in \cite{ciu}, we consider Type IIB string theory configurations involving both D3-branes and RR and NSNS 3-form fluxes. The D3-branes fill-up four non-compact dimensions and sit at a particular point in a six-dimensional manifold ${\bf \cam_6}$, whereas the 3-form fluxes are turned on transversally to the D3-brane worldvolume, i.e., the fluxes have support on 3-forms in ${\bf \cam_6}$ (see Figure \ref{fluxD3}). In this scenario, the open string degrees of freedom associated to the D3-branes yield a supersymmetric gauge theory in their $D=4$ worldvolume. The closed string background surrounding the D3-branes, on the other hand, does not preserve the same supersymmetry as the D3's, and hence induces $\a'$ corrections in the open string sector. From the point of view of the D3's worldvolume theory, these are seen as SUSY-breaking soft terms.

A general derivation of the SUSY-breaking soft terms for D3-branes in a generic background was given in \cite{ciu}, with special emphasis on the case where the source of SUSY breaking is given by the background 3-form fluxes. We will now search for specific models where such general results can be applied. In order to have chiral $\cn = 1$ $D=4$ gauge theories from D3-branes we need to consider a compact manifold ${\bf \cam_6}$ containing singular points, such as orbifold/orientifold singularities, and place the D3-branes on top of them. The low energy spectrum of such gauge theory can be encoded in a quiver diagram \cite{firstquiver}, and can lead to $D=4$ chiral theories quite close to the Standard Model \cite{botup,singu}.

The consistency of these kind of models usually requires the presence of D7-branes at this same singular point \cite{botup}. These new open string sectors will also introduce gauge theories, and possibly extra chiral matter. One could then, in principle, build up a semi-realistic model by using both D3 and D7-branes, as is usually the case in the literature. On the other hand, the computation of soft terms is microscopically better understood in the case of D3-branes. For this reason, we will focus on models where the gauge theory and chiral matter of interest is fully embedded in the $D3-D3$ sector of the model, and D7-branes only play the role of a hidden sector in the low energy theory.

%%%%%%%%%%%%%%%%%%%%%
%%%%%%%%%%%%%%%%%%%%%

We then introduce non-trivial RR ($F_3$) and NSNS ($H_3$) background fluxes which will induce the soft terms in the D3-brane worldvolume theory. These fluxes should be well-defined in the local orbifold geometry which the D3-branes see. The appearance of SUSY-breaking operators of lower dimension is specified by the local behaviour of the flux near the D3-brane location \cite{ciu}. We can then, in principle, consider a local orbifold/orientifold model and perform a local analysis of the flux-induced SUSY-breaking, much in the spirit of \cite{grana}. Indeed, the soft term structure can already be analyzed at this local level, without knowing the specific embedding of the singularity in $\cam_6$.

Eventually, however, we would like ${\bf \cam_6}$ to be compact in order to recover $D=4$ gravity at low energies. The fluxes should then be properly quantized in terms of the homology/cohomology of ${\bf \cam_6}$. Moreover, the whole construction must satisfy the full set of consistency conditions known as RR tadpole cancellation conditions, many of them not visible from the (local) point of view of the D3's gauge theory \cite{local}. As we will see in the next section, these global constraints allow us to study other kind of questions otherwise invisible from the local point of view, such as the distribution of phenomenologically viable models in the space of well-quantized fluxes.

Global models of the type described above, based on toroidal orientifolds with non-trivial background fluxes, have already been built in \cite{cu,ciu}. These constructions are based on D3-branes at $\cpx^3/\inte_3$ orbifold singularities,\footnote{More precisely, these models are based on either D3-branes or anti-D3-branes at orbifold singularities, surrounded by a 3-form flux which breaks supersymmetry. For a more complete discussion of the different possibilities within this scenario see \cite{throat}.} which are the more appealing local models from the phenomenological point of view, since they naturally provide triplication of families \cite{botup}. It turns out, however, that these $\inte_3$ orbifold geometries, and in particular the global $\T^6/\inte_3$ orientifold models constructed in \cite{cu,ciu}, forbid any constant component of the 3-form flux $G_3$ other than the ones proportional to the holomorphic 3-form and its complex conjugate. From the point of view of the general analysis in \cite{ciu}, this will imply that only SUSY-breaking soft terms of the `dilaton domination form' will appear in the effective theory and, e.g., no $\mu$-term would be generated.

In this section we build examples of chiral $\cn=1$ gauge theories based on D3-branes at orbifold/orientifold singularities, which admit a more general pattern of flux-induced soft terms. We will focus on a $\cpx^3/\inte_4$ orbifold singularity, which is the simplest orbifold example allowing for a flux-induced $\mu$-term. Following the general philosophy in \cite{botup}, we first construct the local features of the D3-brane model, and then present an example of how this local physics con be embedded in a global construction.

\subsection{A local $\inte_4$ model}

We now describe in some detail the construction of a local $\inte_4$ D3-brane model in the background of SUSY breaking 3-form fluxes. The aim is to find a model which allows for a semi-realistic gauge group and chiral fermions with appropriate quantum numbers, both arising from strings ending just on D3-branes. This will allow for a more transparent analysis of the effect of flux SUSY breaking, which can already be addressed at the local level. We perform such analysis in Section \ref{Phenomenology}. Although $\inte_4$ does not allow for more than two generation models, the phenomenological analysis of Section \ref{Phenomenology} shows that the soft SUSY breaking terms follow an universal pattern, and so we find that, in order to test the phenomenological possibilities of flux-induced SUSY breaking, the number of generations is not a crucial factor of the model.

In the following we will use the results and conventions in \cite{botup}, to which the reader is referred for more details and a general discussion on model building of D3-branes at singularities.

\subsubsection{Orbifolding}

\begin{figure}[ht]
\centering
\epsfxsize=3in
\epsffile{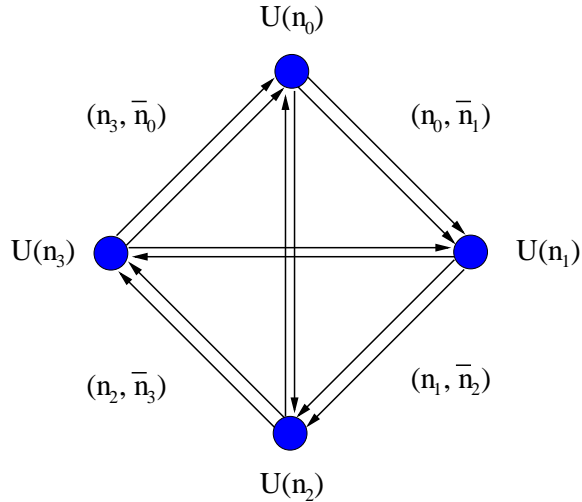}
\caption{\small{Quiver diagram of the $\inte_4$ orbifold. Here the nodes represent $U(n)$ gauge groups and the arrows bifundamental $(n,\bar{n'})$ representations of left-handed chiral multiplets between them, pointing in the sense of the anti-fundamental.}}
\label{quiverZ4}
\end{figure}

Let us consider a stack of $N$ D3-branes sitting on top of a supersymmetric $\inte_4$ orbifold singularity. Locally, we can represent such singularity by $\cpx^3/\inte_4$, where the generator of the $\inte_4$ orbifold action $\th$ acts geometrically on $\cpx^3$ as
\beq
\th :\ (z_1,z_2,z_3)\ \mapsto\ (e^{-2\pi i/ 4} z_1, e^{-2\pi i/ 4} z_2, e^{2\pi i/ 2} z_3)\ = \ (-i z_1, -iz_2, -z_3)
\label{geomZ4}
\eeq
and is encoded in the shift vector $v = \frac{1}{4} (1,1,-2)$.\footnote{We have introduced the shift vector $v$ in order to simplify the discussion. In the language of \cite{botup}, their components $v_r$ correspond to $a_r/N$.} On the other hand, the action on the D3's Chan-Paton degrees of freedom is given by
\beq
\g_{\th,3} = {\rm diag\ } \left(\id_{n_0}, i \id_{n_1}, - \id_{n_2}, -i \id_{n_3} \right) 
\label{ChanZ4}
\eeq
with $\sum_{i=1}^4 n_i = N$. The massless spectrum of open strings starting and ending on D3-branes can be obtained by considering an $N \times N$ matrix $\l_{33}$ and imposing, on the Neveu-Schwarz sector, the conditions
\beqa \label{gaugeD3}
\l_{33}^{(0)} = \g_{\th,3} \l_{33}^{(0)} \g_{\th,3}^{-1} & {\rm for} & \l_{33}^{(0)} \psi_{-\oh}^\mu \vac \\
\l_{33}^{(r)} = e^{2\pi i v_r} \g_{\th,3} \l_{33}^{(r)} \g_{\th,3}^{-1} & {\rm for} & \l_{33}^{(r)} \psi_{-\oh}^r \vac
\label{chiralD3}
\eeqa
where $\mu$ labels the D3-brane world-volume dimensions, (\ref{gaugeD3}) representing $D=4$ gauge bosons, and $r = 1, 2, 3$ labels each of the complex scalars coming from each complex plane in $\cpx^3$. The conditions to be imposed in the Ramond sector are the same, the orbifold being supersymmetric, and we obtain a massless spectrum completely specified by the four integers $\{n_i\}_{i=1}^4$ 
\beq
\begin{array}{ll}\vspace*{.2cm}
{\rm \bf Vector\ Multiplets} & \prod_{i=1}^{4} U(n_i) \\
{\rm \bf Chiral\ Multiplets} & \sum_{r=1}^{3} \sum_{i=1}^4 (n_i, \bar{n}_{i+4 v_r})
\end{array}
\label{specD3}
\eeq
which the index $i$ is to be understood mod 4. 
This spectrum can be encoded in the quiver diagram of Figure \ref{quiverZ4}. This quiver construction can be generalized to arbitrary orbifold singularities, see e.g. \cite{quivers}.
%
%%%%%%%%%%%%%%%%%%%%%
%%%%%%%%%%%%%%%%%%%%%

The quiver diagram shows clearly how the orbifold singularity $\inte_4$ provides a chiral spectrum. Moreover, the chiral multiplets $(n_i, \bar{n}_{i+1})$ come in pairs, which implies two replicas of the same kind of particle. Of course, from a phenomenological point of view the natural triplication of families is far more appealing, which can be obtained, e.g., by means of $\inte_3$ orbifolds \cite{botup}. However, as explained above, these orbifolds impose too many constraints on the background fluxes.

In general we can also consider D7-branes filling two complex dimensions in $\cpx^3/\inte_4$ and fixed by the geometrical orbifold action. Let us, for instance, consider D7-branes not wrapping the 3$^{rd}$ complex plane, denoted by D7$_3$-branes, with the Chan-Paton action of the orbifold specified by
\beq
\g_{\th,7_3} = {\rm diag\ } \left(\id_{u_0}, i \id_{u_1}, - \id_{u_2}, -i \id_{u_3} \right) 
\label{ChanZ4D7}
\eeq
The quiver structure of Figure \ref{quiverZ4} equally applies to this case. However, since the `internal' volume of the D7-branes is infinite, the gauge coupling constants of the corresponding gauge groups vanish and they become global symmetries of the theory.\footnote{Of course, in a compact model we recover such gauge symmetries.} The matter charged under such global sector and the D3's gauge groups are open strings in the $37_3$ and $7_33$ sectors. The orbifold projection is now
\beq
\l_{37_3} = e^{-i \pi v_3} \g_{\th,3} \l_{37_3} \g_{\th,7_3}^{-1} \quad \quad \l_{7_33} = e^{-i \pi v_3} \g_{\th,7_3} \l_{7_33} \g_{\th,3}^{-1}
\label{spec73}
\eeq
yielding a chiral spectrum which, in the case of $\inte_4$ is
\beq
\begin{array}{ll}
{\rm \bf Chiral\ Multiplets} & \sum_{i=1}^4 [(n_i, \bar{u}_{i+1}) + (\bar{n}_{i}, u_{i-1})]
\end{array}
\label{specD3D7}
\eeq

Finally, for consistency of the theory tadpoles must be cancelled. A local model is only sensitive to twisted tadpole conditions, which in the case of D3/D7-branes at orbifold singularities read \cite{botup}
\beq
\prod_{r=1}^3 2 {\rm sin}(\pi k v_r) \Tr \g_{\th^k,3} + \sum_{r=1}^3 2 {\rm sin} (\pi k v_r) \Tr \g_{\th^k,7_r} = 0
\label{twisted}
\eeq

In the case at hand we have
\beq
\begin{array}{ccc}
k=1 & & 2 \Tr \g_{\th^1,3} + \Tr \g_{\th^1,7_3} = 0 \\
k=2 & & \left[ 4 \Tr \g_{\th^2,3} + \Tr \g_{\th^2,7_3} = 0 \right] \\
\end{array}
\label{twistedZ4}
\eeq
where $\g_{\th^2,3} = \g_{\th^1,3}^2$, etc. Notice that the case $k=0$, corresponding to untwisted tadpoles, identically vanishes. This is because the RR and NSNS tadpoles are suppressed by the volume transverse to the D-brane where the closed string fields carrying the corresponding D-brane charge can propagate. In the case of untwisted charges the RR (or NSNS) fields propagate all over $\cpx^3/\inte_4$, which is infinite volume. Actually, the same is true for the twisted tadpole $k=2$, since the rotation $\th^2$ leaves the third complex plane fixed, so the corresponding twisted closed strings are not stuck at the orbifold singularity and can propagate all over this complex plane. As a result, the $k=2$ condition above need not be taken into account in order to build a local $\inte_4$ orbifold model.

To summarize, from the point of view of the local physics at the singularity, and hence the gauge field theory physics, the only relevant consistency condition is the twisted tadpole for $k=1$. Indeed, cancellation of chiral anomalies will be guaranteed by fulfilling such consistency condition \cite{local}. Thus, at this level of the construction, we only need to bother about $k=1$ twisted conditions. However, untwisted and $k=2$ twisted conditions will be relevant when embedding such singularity in a compact model, as we will show below.

Given this general spectrum we can think of building a semi-realistic local model with two generations of quarks and leptons. To obtain Standard-like or Pati-Salam models, the more appealing choices are
\beq
\begin{array}{cccc}
n_0 = 3 & n_1 = 2 & n_2 = 1 & n_3 = 1 \\
n_0 = 3 & n_1 = 2 & n_2 = 2 & n_3 = 1 \\
n_0 = 4 & n_1 = 2 & n_2 = 1 & n_3 = 0 \\
n_0 = 4 & n_1 = 2 & n_2 = 2 & n_3 = 0 \\
\end{array}
\label{trial}
\eeq
all of these choices lead, however, to uncancelled $k=1$ tadpoles that will require the presence of D$7_3$-branes and will, eventually, lead to unwanted extra chiral matter in the model. As pointed out in \cite{imr} this non-minimality of the spectrum is generic in orbifold models. We thus turn to consider orientifolded versions of the above.

\subsubsection{Orientifolding}

\begin{figure}[ht]
\centering
\epsfxsize=4.5in
\epsffile{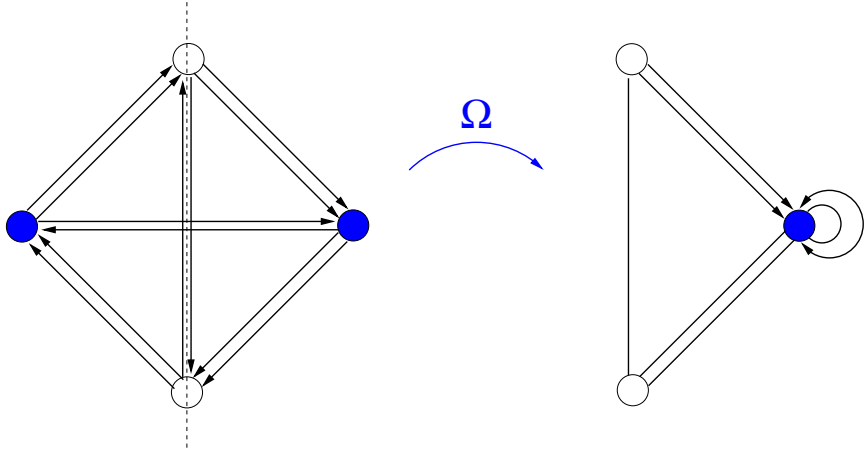}
\caption{\small{Folding of the $\inte_4$ quiver diagram in Figure \ref{quiverZ4} to the corresponding orientifold with vector structure. The colored nodes represent pairs of $U(n)$ gauge groups identified under the action of $\Om \cR$, while the white nodes are $U(m)$ gauge groups projected to either $SO(m)$ or $USp(m)$.}}
\label{folding}
\end{figure}

Let us now consider type IIB string theory modded out by $\Om \cR$, where $\cR$ is a $\inte_2$ symmetry of the compactification manifold $\cam_6$. Such class of theories are indeed a natural way to achieve non-trivial supersymmetric flux compactifications \cite{drs,gkp}. Let us, moreover, consider that our previous $\inte_4$ orbifold singularity is also fixed by the action of $\cR$. In our local model $\cpx^3/\inte_4$, we can take $\cR: z_i \mapsto - z_i$.\footnote{More precisely, we are considering the standard orientifold projection leading to O3-planes and preserving $\cn =1$ supersymmetry, which is of the form $\Om \cR$, with $\cR = R_1 R_2 R_3 (-1)^{F_L}$. Here $R_i$ acts as $z_i \mapsto - z_i$, and $F_L$ is the left-handed world-sheet fermion number.} Since the D3-branes are at a fixed point of the new orientifold symmetry, we should project the open string degrees of freedom to those invariant both under $\inte_4$ and $\Om \cR$, in order to get the final gauge theory. The conditions (\ref{gaugeD3}) and (\ref{chiralD3}) now become
\beqa \label{gaugeD3o}
\l^{(0)} = \g_{\th,3} \l^{(0)} \g_{\th,3}^{-1} & \l^{(0)} = - \g_{\Om,3} {\l^{(0)}}^T \g_{\Om,3}^{-1} \\
\l^{(r)} = e^{2\pi i v_r} \g_{\th,3} \l^{(r)} \g_{\th,3}^{-1} & \l^{(r)} = - \g_{\Om,3} {\l^{(r)}}^T \g_{\Om,3}^{-1}
\label{chiralD3o}
\eeqa
where $\g_{\Om,3}$ is the action of $\Om \cR$ on the Chan-Paton degrees of freedom. Again, $\l^{(0)}$ corresponds to the gauge bosons and $\l^{(r)}$ to each complex scalar. String interactions are well-defined if both conditions 
\beq
\begin{array}{c}\vspace*{.2cm}
\g_{\Om}^T \g_{\Om}^{-1} = \pm \id,\\
\g_{\th^k}^* = \pm \g_{\Om}^* \g_{\th^k} \g_{\Om}
\end{array}
\label{consistency}
\eeq
are satisfied by the unitary matrices $\g_{\Om}$, $\g_{\th^k}$, which restricts the possible form of $\g_{\Om}$ to a few cases \cite{tensors}. In general, the orientifold symmetry will identify pairs of gauge groups and chiral matter multiplets which were present in the orbifold theory, while some other unitary gauge groups may be reduced to $SO$ or $USp$ gauge groups. Such action can be again encoded in a `folding' of the orbifold diagram, as shown in Figure \ref{folding} for the particular case of interest in this paper.
%
%%%%%%%%%%%%%%%%%%%%%
%%%%%%%%%%%%%%%%%%%%%

The procedure of obtaining orientifold from orbifold quivers can be summarized in a set of rules (for a general discussion of these rules, see, e.g., \cite{raul}). In this paper, we will be interested in the orientifold projection first discussed in \cite{tensors}, and which in the closed string sector gives rise to left-right symmetric RR states from the $\inte_2$ twisted sectors of the theory. In the particular case of $\inte_4$ such projection imposes the constraint
\beq
\g_\Om = \g_{\th^2} \g_\Om \g_{\th^2}^T
\label{constraint}
\eeq
for both D3 and D7-branes. Together with (\ref{consistency}) this implies $\g_{\th,3}^4 = \g_{\th,7}^4 = \id$, which is known as an orientifold with vector structure.\footnote{At this stage, one may also consider $\inte_4$ orientifolds without vector structure. However, these orientifolds possess exotic twisted tadpole conditions \cite{afiv}, which usually imply the presence of non-BPS D-branes in order to find a consistent vacuum \cite{raulangel}. This would mean an extra source of supersymmetry breaking, and thus does not seem very promising for our purposes.} The two inequivalent choices of $\g_{\Om}$ are given by
\beq
\g_{\Om,a} =
\left(
\begin{array}{cccc}
\id_{n_0} \\
& & & \id_{n_1} \\
& & \id_{n_2} \\
& \id_{n_3}
\end{array}
\right)
\quad
\g_{\Om,b} =
\left(
\begin{array}{cccc}
\eps_{n_0} \\
& & & \id_{n_1} \\
& & \eps_{n_2} \\
& \id_{n_3}
\end{array}
\right)
\label{Chanori}
\eeq
where $\eps_n$ is a block-diagonal matrix with $n$ blocks of the form 
{\footnotesize $\left(
\begin{array}{cc}
0 & 1 \\
-1 & 0 
\end{array}
\right)$}. Moreover, in order for $\g_{\Om}$ to have a well-defined action on the Chan-Paton matrices, we must impose $n_1 = n_3$, $u_1 = u_3$ in (\ref{ChanZ4}) and (\ref{ChanZ4D7}), respectively.

One choice of $\g_{\Om}$ or the other will depend on the RR charges of the O-planes placed at the orientifold singularity. More specifically, an O3-plane with NSNS and RR untwisted charges opposite to those of a D3-brane, denoted as $O3^{(-,-)}$, will correspond to the choice $\g_{\Om,3} = \g_{\Om,a}$, while that with the same charges, $O3^{(+,+)}$, corresponds to $\g_{\Om,3} = \g_{\Om,b}$. Notice that the orientifold group element $\Om \cR \th^2$ fixes the two first complex planes of $\cpx^3$, and hence also O7-planes will be present in our local model. Consistency conditions imply that the charges of the O7-plane are opposite to those of the companion O3-plane. We hence have the two possibilities $(O3^{(-,-)},O7^{(+,+)})$, which implies $\g_{\Om,3} = \g_{\Om,a}$ and $\g_{\Om,7_3} = \g_{\Om,b}$, and $(O3^{(+,+)},O7^{(-,-)})$, which corresponds to the opposite choice. Finally, in the particular case of $\inte_4$ such O-planes do not have any twisted charges, which implies that the tadpole conditions (\ref{twistedZ4}) remain unchanged.

Of course, the D3-brane gauge theory and chiral spectrum will strongly depend on the choice of $\g_{\Om,3}$, so let us discuss both possible choices (\ref{Chanori}) in turn.

\begin{itemize}

\item{\bf  $O^{(-,-)}$}

In this case the extra projection in (\ref{gaugeD3o}) identifies the gauge groups $U(n_1)$ and $U(n_3)$ (recall that $n_1 = n_3$) and projects $U(n_0)$ and $U(n_2)$ into $SO(n_0)$ and $SO(n_2)$, respectively. The $\cn=1$ chiral multiplets are summarized in table \ref{minus} and Figure \ref{quiverZ4o-}.
%
%%%%%%%%%%%%%%%%%%%%%
\begin{figure}[!htb]
\centering
\epsfxsize=3in
\epsffile{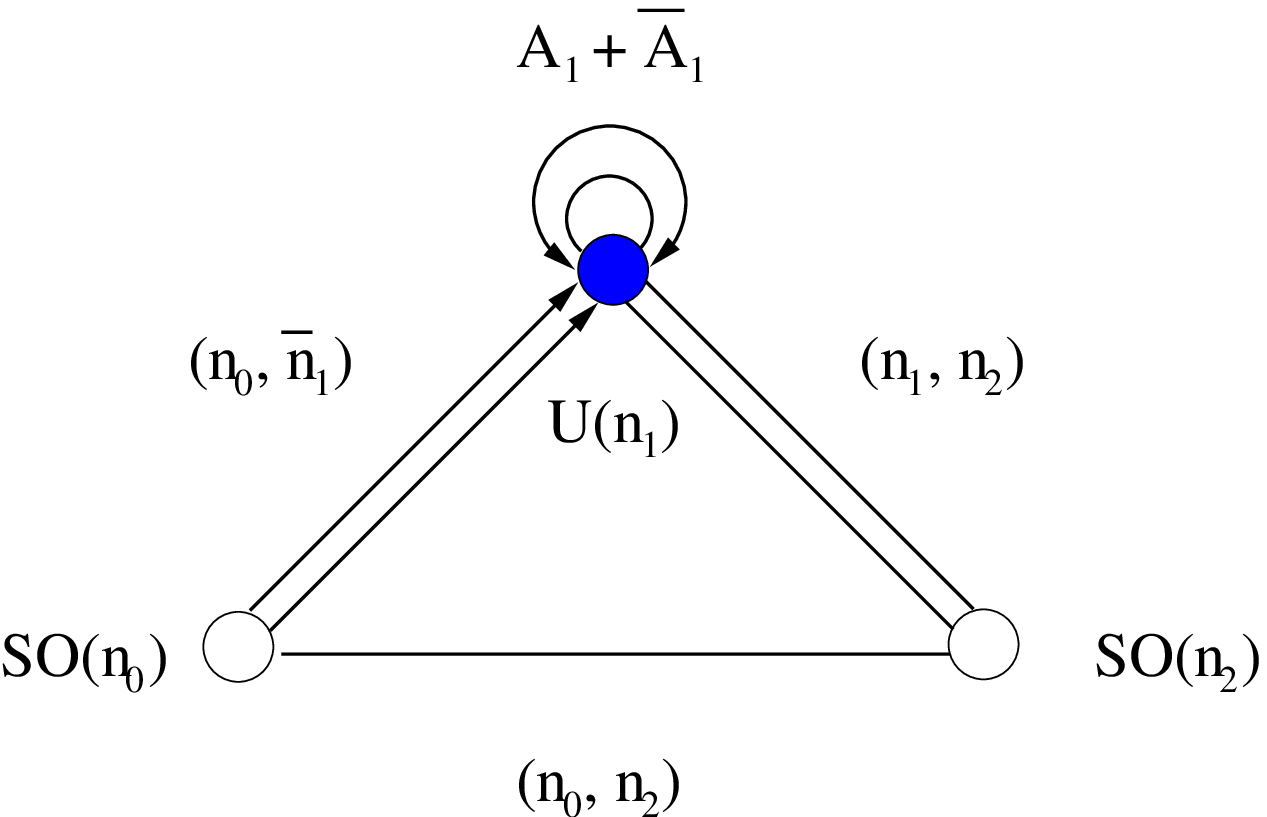}
\caption{\small{Quiver diagram of the $\inte_4$ orientifold with the choice $\g_{\Om,a}$.}}
\label{quiverZ4o-}
\end{figure}
%%%%%%%%%%%%%%%%%%%%%
%
\begin{table}[htb]
\renewcommand{\arraystretch}{1.25}
\begin{center}
\begin{tabular}{cc}
%\hline
{\bf Gauge group} & $SO(n_0) \times U(n_1) \times SO(n_2)$ \\
{\bf Chiral multiplets} & $2 (n_0, \bar{n}_1)$, $2 (n_1, n_2)$, $(n_0, n_2)$, ${\bf A_1} + {\bf \bar{A}_1}$ \\
%\hline
\end{tabular}
\caption{\small Spectrum of the $\inte_4$ orientifold with the choice $\g_{\Om,a}$. Here ${\bf A_1}$ stands for the antisymmetric representation of $U(n_1)$.
\label{minus}}
\end{center}
\end{table}

\item{\bf  $O^{(+,+)}$}

This projection again identifies the unitary gauge groups $U(n_1)$ and $U(n_3)$, while $U(n_0)$ and $U(n_2)$ are now projected to $USp$ groups. Notice that such group only makes sense if both $n_0$ and $n_2$ are even integers. We again summarize the $\cn=1$ gauge group and chiral spectrum in table \ref{plus} and Figure \ref{quiverZ4o+}.
%
%%%%%%%%%%%%%%%%%%%%%
\begin{figure}[!htb]
\centering
\epsfxsize=3in
\epsffile{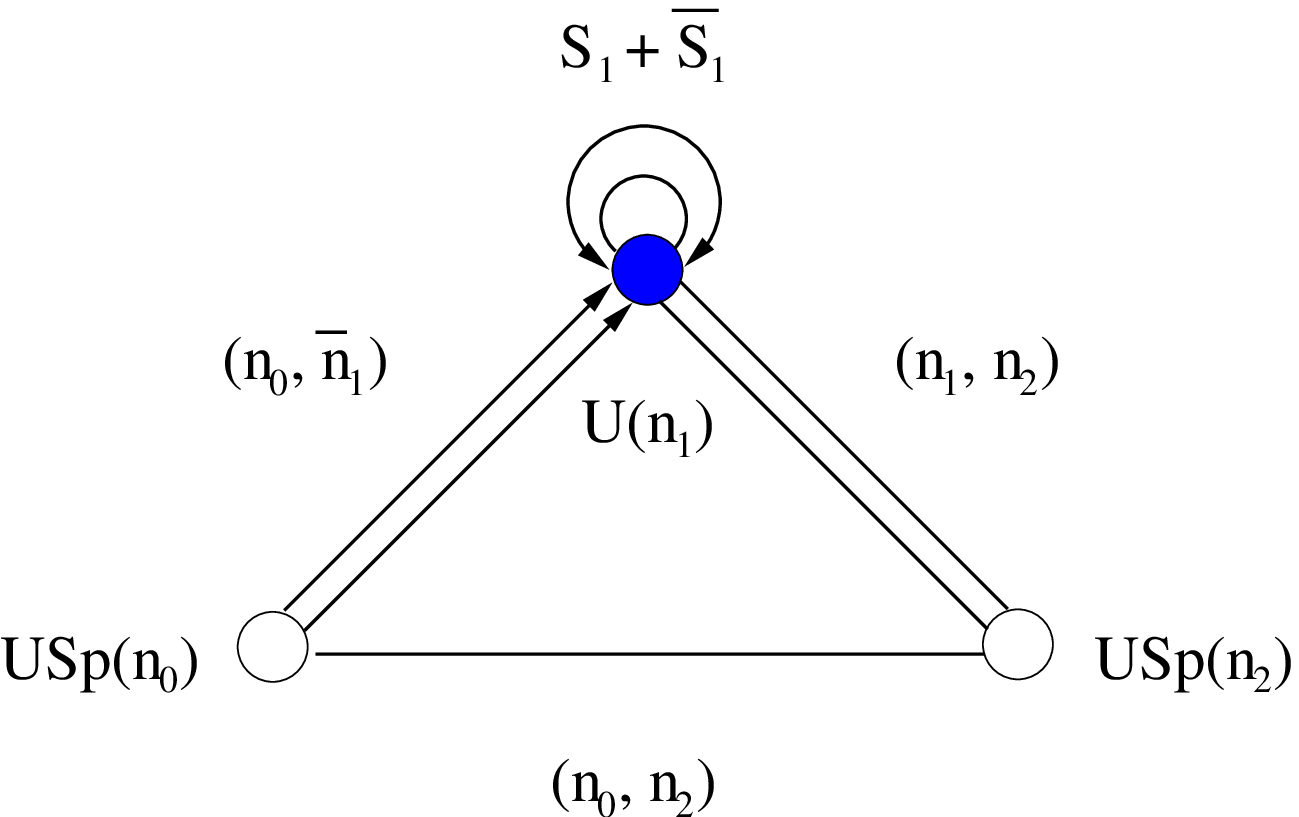}
\caption{\small{Quiver diagram of the $\inte_4$ orientifold with the choice $\g_{\Om,b}$.}}
\label{quiverZ4o+}
\end{figure}
%%%%%%%%%%%%%%%%%%%%%
%
\begin{table}[htb]
\renewcommand{\arraystretch}{1.25}
\begin{center}
\begin{tabular}{cc}
%\hline
{\bf Gauge group} & $USp(n_0) \times U(n_1) \times USp(n_2)$ \\
{\bf Chiral multiplets} & $2 (n_0, \bar{n}_1)$, $2 (n_1, n_2)$, $(n_0, n_2)$, ${\bf S_1} + {\bf \bar{S}_1}$ \\
%\hline
\end{tabular}
\caption{\small Spectrum of the $\inte_4$ orientifold with the choice $\g_{\Om,b}$. Here ${\bf S_1}$ stands for the symmetric representation of $U(n_1)$.
\label{plus}}
\end{center}
\end{table}

\end{itemize}

These spectra apply to D7-branes as well, by just making the substitution $n_i \mapsto u_i$. More specifically, if we choose the projection $(O3^{(-,-)},O7_3^{(+,+)})$ the D3-brane spectrum will be given by Figure \ref{quiverZ4o-} and D7-brane spectrum by Figure \ref{quiverZ4o+}, and the other way round if we choose $(O3^{(-,-)},O7_3^{(+,+)})$. Finally, the orientifold projection identifies the $37_3$ and $7_33$ sectors, so in order to compute the $D3-D7_3$ spectrum we only need to consider one of the contributions in (\ref{spec73}). The final result is given in table \ref{mixed}.
\begin{table}[htb]
\renewcommand{\arraystretch}{1.25}
\begin{center}
\begin{tabular}{cc}
%\hline
{\bf Chiral multiplets} & $(n_0, \bar{u}_1)$, $(n_1, u_2)$, $(n_2, u_1)$, $(\bar{n}_1, u_0)$ \\
%\hline
\end{tabular}
\caption{\small Spectrum of the $\inte_4$ orientifold in the mixed $37_3 + 7_33$ sector.
\label{mixed}}
\end{center}
\end{table}

\subsubsection{Massive $U(1)$'s}

Before any attempt to build a model, it is important to know which of the above $U(1)$'s are actual gauge symmetries in the low energy theory. Indeed, given a configuration of D3-branes at singularities the general gauge group will contain $U(n) \simeq SU(n) \times U(1)$ factors and many of these $U(1)$'s can be seen to develop mixed $U(1)$-non Abelian gauge anomalies from the chiral spectrum of the theory. Such field theory anomalies are cancelled by a generalized Green-Schwarz mechanism mediated by closed string twisted modes \cite{iru}. As a result, these $U(1)$ gauge bosons get a Stueckelberg mass, remaining in the effective field theory as global symmetries.

Let us first consider the orbifold spectrum (\ref{specD3}). By computing the field theory $U(1)_a-SU(n_b)^2$ anomalies, $a,b = 0, \dots, 3$, is easy to see that the only non-anomalous $U(1)$'s are given by
\beq
\begin{array}{lcr}\vspace*{.1cm}
Q_1 & = & \frac{Q_0}{n_0} + \frac{Q_2}{n_2} \\
Q_2 & = & \frac{Q_1}{n_1} + \frac{Q_3}{n_3} \\
\end{array}
\label{nonanorbi}
\eeq
where $Q_i$ corresponds to the generator of the $i^{th}$ $U(1)$. Notice that $Q_{\rm diag} = Q_1 + Q_2$ is the diagonal $U(1)$ which is always non-anomalous in this class of models \cite{botup}.\footnote{This is generically the only non-anomalous $U(1)$. In the particular case of $\inte_4$ we have an extra one, $Q_1 - Q_2$, basically because $\th^2$ leaves one complex plane fixed.}

On the other hand, when consider an orientifold model, there is essentially one $U(1)$ factor, now inside $U(n_1)$, which can be seen to be anomalous. The only way to get a non-anomalous Abelian gauge factor seems to be to consider $SO(2) \simeq U(1)$ gauge groups from Figure \ref{quiverZ4o-}.

\subsubsection{Constructing the local model}

Let us now focus on the model building possibilities of the $\inte_4$ orientifolds described above. Since the Abelian factor of $U(n_1) \simeq SU(n_1) \times U(1)$ is anomalous (and hence does not survive as a gauge symmetry at low energies), it turns difficult to find a natural candidate for the hypercharge or $U(1)_{B-L}$ characteristic of Left-Right symmetric models. We will then consider building Pati-Salam models, which do not include any Abelian gauge factor before gauge symmetry breaking. If we consider the choice of orientifold projection $(O3^{(-,-)},O7^{(+,+)})$, the most realistic gauge groups that we can achieve in the D3-brane worldvolume theory are given by the choices
\beq
\begin{array}{cccc}
n_0 = 3 & n_1 = 4 & n_2 = 3 \\
n_0 = 2 & n_1 = 4 & n_2 = 3 & \Raw \quad u_0 - u_2 = 2
\end{array}
\label{trialO-}
\eeq
where in the second choice we have imposed the twisted tadpole condition (\ref{twistedZ4}) for the twist $k=1$. Although these choices may lead to semi-realistic gauge groups by using the identifications $SO(3) \simeq SU(2)$ and $SO(2) \simeq U(1)$, notice that the would-be Higgs and quarks are triplets instead of doublets of $SO(3)$.

\begin{figure}[!htb]
\centering
\epsfxsize=3in
\epsffile{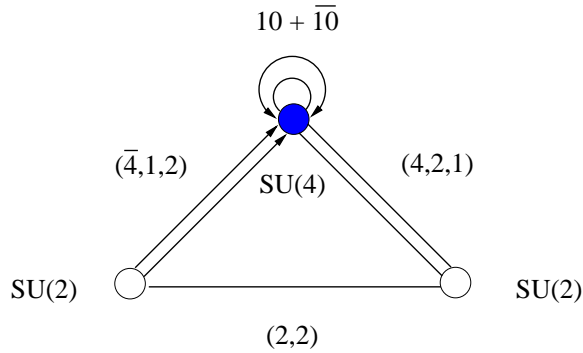}
\caption{\small{Pati-Salam $\inte_4$ orientifold model with two-generations.}}
\label{quiverZ4PS}
\end{figure}

A more appealing possibility is to consider the orientifold projection $(O3^{(+,+)},O7^{(-,-)})$. Indeed, we can now make use of the identity $USp(2) \simeq SU(2)$ and achieve a Pati-Salam gauge group by the choice
\beq
\begin{array}{ccc}
n_0 = 2 & n_1 = 4 & n_2 = 2
\end{array}
\label{trialO+}
\eeq
as shown in Figure \ref{quiverZ4PS}.
%
%%%%%%%%%%%%%%%%%%%%%
%%%%%%%%%%%%%%%%%%%%%

Notice that the $k=1$ twisted tadpole condition in (\ref{twistedZ4}) is satisfied, so that in principle we do not need the presence of D7-branes in our local model.
 Notice as well that the chiral spectrum of this $\cn =1$ Pati-Salam model contains two replicas of chiral matter and only one Higgs multiplet.

\subsection{A global $\inte_4$ model}

Let us now construct an example of a fully-fledged string compactification where the $\inte_4$ orientifold model of Figure \ref{quiverZ4PS} can be embedded. In such construction we will also include the background fluxes that will induce the soft terms in the supersymmetric D3-brane effective field theory. Considering a particular global model does not only specify the gravity sector of the theory as well as possible open string hidden sectors but, as shown below, it also imposes some global constraints on the background fluxes. As shown in Section \ref{Phenomenology}, these constraints may be relevant when considering certain phenomenological aspects such as Electroweak Symmetry Breaking in this context.
%
%%%%%%%%%%%%%%%%%%%%%%%%
\begin{figure}[!htb]
\centering
\epsfxsize=5in
\hspace*{0in}\vspace*{.2in}
\epsffile{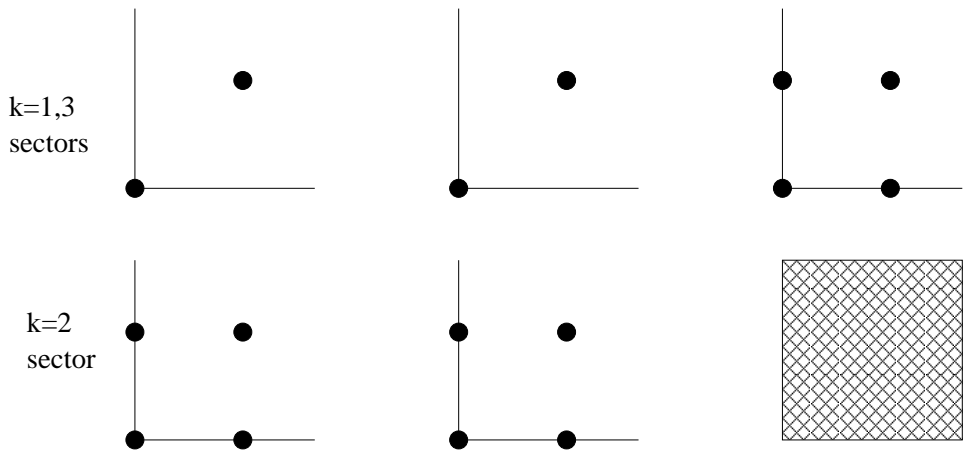}
\caption{\small Representation of the fixed sets of the $\T^6/\inte_4$ orbifold. The three squares represent the three $\T^2$ tori. Dots are fixed points in the torus and a grid in a square corresponds to a fixed torus. 
%Figure taken from \cite{raul}.} 
}
\label{T6Z4} 
\end{figure}
%%%%%%%%%%%%%%%%%%%%%%%%%
%

The simplest compact model where $\cpx^3/\inte_4$ singularities arise is the toroidal orbifold $\T^6/\inte_4$, shown in Figure \ref{T6Z4}. We are interested in orientifolds of such singular space containing $O3^{(+,+)}$'s. This can be achieved by modding the theory by $\Om \cR$, where $\cR$ contains the $\inte_2$ geometrical action $(z_1, z_2, z_3) \mapsto (-z_1,-z_2,-z_3)$ and satisfies the condition (\ref{constraint}). Such class of models (or rather T-dual versions) have been considered in \cite{ua,aadds}, where anti-D-branes were introduced in order to cancel RR tadpoles.\footnote{We are not particularly interested in introducing anti-D-branes at an orientifold point, since their effective theory is already non-supersymmetric.} In particular, we will consider a T-dual version of the Type I $\inte_4$ model in \cite{aadds}, obtained by T-dualizing in the two directions of the third $\T^2$. Moreover, we will consider the possibility of adding a discrete B-field $b=0,1/2$ in the third $\T^2$ before T-dualizing. After T-dualizing we obtain a model with either 64 $O3^{(+,+)}$'s ($b=0$) or 48 $O3^{(+,+)}$ and 16 $O3^{(-,-)}$ ($b=1/2$) \cite{vector,morevector}. The model will also contain O$7_3$-planes, namely $4 \cdot O7_3^{(-,-)}$ for the choice $b=0$ and $3 \cdot O7_3^{(-,-)}$ + $1 \cdot O7_3^{(+,+)}$ for $b=1/2$. Both O3 and O$7_3$-planes contribute only to untwisted tadpoles with (NSNS,RR) charges
\beq
O3's \raw \left(+32\cdot (1-b), +32 \cdot (1-b) \right), \quad O7_3's \raw \left(-32\cdot (1-b), -32 \cdot 
(1-b) \right)
\label{untwisted}
\eeq
in units of D3 and D7-brane charges, respectively.

Let us now embed the local model described in the previous section in this setup. More precisely, let us consider the D3-brane content (\ref{trialO+}) on top of the origin, which is fixed both under $\inte_4$ and $\Om \cR$. As we showed previously, the twisted tadpole condition for $k=1$ is automatically satisfied, without considering any extra D3-branes or D7-branes. In this compact model, however, we also have to satisfy the $k=2$ twisted tadpole in (\ref{twistedZ4}). This can be easily done by adding a stack of D3'-branes with Chan-Paton factors
\beq
\g_{\th,3'} = {\rm diag\ } \left(\id_{R+2}, i \id_{R}, - \id_{R+2}, -i \id_{R} \right) 
\label{extraD3}
\eeq
with $R \in {\bf N}$. This stack must be placed in an orbifold singularity of the form $(0,0,P)$, where $P = 0, 1/2, i/2, (1+i)/2$ is any of the $\inte_2$ fixed points in the third $\T^2$ (see fig.(\ref{T6Z4})). This is the case because the twisted RR field in the $k=2$ sector propagates over the whole third $\T^2$, and hence the corresponding RR tadpole is blind to positions on the third two-torus.

Let us choose $P=(1+i)/2$. That is, the extra stack of D3-branes is localized on the point $(0,0,(1+i)/2)$. Hence, they can be seen as a hidden sector with respect to the PS-like model at the origin, communicating to the latter only by means of gravitational effects. Similarly, we can introduce a hidden sector of D7$_3$-branes, wrapping the first and second two-tori and localized at $P=(1+i)/2$ in the third $\T^2$. We choose these D7's to have Chan-Paton factors
\beq
\g_{\th,7_3} = {\rm diag\ } \left(\id_{S}, i \id_{S}, - \id_{S}, -i \id_{S} \right) 
\label{solD7}
\eeq
where $S \in {\bf N}$. Since these D-branes are in the regular representation of $\inte_4$, they do not generate any further twisted RR tadpoles.

The spectrum of this hidden sector will depend on the choice of discrete B-field on the third $\T^2$. In both cases $(0,0,(1+i)/2)$ will be a $\inte_4$ orientifold singularity, but the O-plane content will change for the choice of $b$ above. If we choose $b=0$, then the O-plane content at this point is of the form $(O3^{(+,+)},O7^{(-,-)})$, whereas if we choose  $b=1/2$ it switches to $(O3^{(-,-)},O7^{(+,+)})$. The gauge group in both cases is given by
\beqa\nonumber
b = 0 & \Raw &
\left\{
\begin{array}{rcl}\vspace*{.1cm}
{\rm D3'-branes} & \raw & SU(R) \times USp(R+2) \times USp(R+2) \\
{\rm D7_3-branes} & \raw & SU(S) \times SO(S) \times SO(S)
\end{array}
\right.
\\ \nonumber
b = \oh  & \Raw & 
\left\{
\begin{array}{rcl}\vspace*{.1cm}
{\rm D3'-branes} & \raw & SU(R) \times SO(R+2) \times SO(R+2) \\
{\rm D7_3-branes} & \raw & SU(S) \times USp(S) \times USp(S)
\end{array}
\right.
\label{completing}
\eeqa

Since we are varying the charge of some orientifold planes by introducing a discrete B-field, cancellation of untwisted tadpoles does also depends on the choice of $b$. In particular, cancellation of RR D7-brane tadpoles reads
\beq
4 \cdot S = 32 \cdot (1-b)
\label{D7RR}
\eeq
so we must consider $S = 8$ for $b=0$ and $S = 4$ for $b=1/2$. 

\subsubsection{Introducing the flux}

As the final stage of building the model, let us introduce the background 3-form flux, which will cancel the untwisted D3-brane RR-charges, stabilize the dilaton/complex structure moduli and induce soft terms in the D3-brane worldvolume. In general we will have both non-trivial RR and NSNS 3-form field-strength backgrounds, denoted as $F_3$ and $H_3$, respectively. These fluxes must obey the Bianchi identities $dF_3 = dH_3 = 0$, as well as be properly quantized with respect to any 3-cycle $\Sigma \subset \cam_6$ \cite{drs,gkp}
\beq
{1 \over 2 \pi \a^\prime} \int_{\Sigma} F_3 \in 2\pi \inte, \quad \quad
{1 \over 2 \pi \a^\prime} \int_{\Sigma} H_3 \in 2\pi \inte
\label{quant2}
\eeq
Actually, in toroidal orientifold models there is an extra subtlety in flux quantization \cite{fp}, which imposes the integer number in (\ref{quant2}) to be even whenever $\Sigma$ is passing through an even number of $O3^{(+,+)}$. This will be the case in the $\T^6/\inte_4$ orientifold model discussed above.

The two real 3-form fluxes are usually arranged into a complex 3-form flux as
\beq
G_3 = F_3 - \tau_4 H_3
\label{cpxflux}
\eeq
where $\tau_4 = a + i/g_s$ is the usual type IIB axion-dilaton coupling. This field-strength flux is a source of the self-dual five-form, hence carries D3-brane RR and NSNS charge. The RR charge is given by the topological number
\beq
Q_{\rm flux} = {1 \over (4\pi^2 \a^\prime)^2} \int_{\cam_6} H_3 \wedge F_3 = {1 \over (4\pi^2 \a^\prime)^2} {i \over 2\pim \phi} \int_{\cam_6}  {G_3 \wedge \overline{G}_3 \over 2\pim \tau_4}
\label{RRcharge-}
\eeq
which is an integer number given the quantization conditions (\ref{quant2}). The sign of this RR charge depends on the particular decomposition of $G_3$ in imaginary self-dual (ISD) and imaginary anti-self-dual (IASD) components. In the particular case of $\T^6/\inte_4$, there are four (untwisted) 3-forms surviving the orbifold projection. They are listed in table \ref{3forms}, with their decomposition in holomorphic indexes and their Hodge duality transformations.
\begin{table}[htb]
\renewcommand{\arraystretch}{1.25}
\begin{center}
\begin{tabular}{|c|c|c|}
\hline
& 3-forms & $*_6$ \\
\hline
$\begin{array}{c}
(0,3) \\ (2,1)
\end{array}$ &
$\begin{array}{c}
d\bar{z}_1 \wedge d\bar{z}_2 \wedge d\bar{z}_3 \\
dz_1 \wedge dz_2 \wedge d\bar{z}_3
\end{array}$
& ISD \\
\hline
$\begin{array}{c}
(3,0) \\ (1,2)
\end{array}$ &
$\begin{array}{c}
dz_1 \wedge dz_2 \wedge dz_3 \\
d\bar{z}_1 \wedge d\bar{z}_2 \wedge dz_3
\end{array}$
& IASD \\
\hline
\end{tabular}
\caption{\small Untwisted cohomology of $\T^6/\inte_4$. ISD stands for the 3-forms $\om$ satisfying $*_6 \om = i \om$, and IASD for those satisfying $*_6 \om = -i \om$.
\label{3forms}}
\end{center}
\end{table}

In the following we will consider our flux $G_3$ to be a linear combination of the IASD 3-forms in table \ref{3forms}, which implies that they will carry negative RR D3-brane charge. Moreover, this will guarantee that the scalar potential generated by $G_3$ will be at its minimum, freezing the complex structure moduli $\tau_i$ $i=1,2,3$ of $\T^2 \times \T^2 \times \T^2$ and the complex dilaton $\tau_4$ to some specific values.\footnote{Actually, the $\inte_4$ geometrical action already fixes the first two $\T^2$'s to be square tori, so we have $\tau_1 = \tau_2 = i$ and $\tau_3$, $\tau_4$ as free parameters.} More specifically, we will take the 3-form flux to be of the form
\beq
G_3\ =\ A \cdot dz_1 \wedge dz_2 \wedge dz_3\ +\ B \cdot d\bar{z}_1 \wedge d\bar{z}_2 \wedge dz_3
\label{G3}
\eeq
with $A$, $B \in \cpx$. There are two equivalent approaches in the literature for finding flux vacua. One of them is considering some well quantized fluxes $F_3$ and $H_3$ and looking for points in the moduli space of complex structures $+$ dilaton where the scalar potential vanishes \cite{kst}. The other is fixing $G_3$ to be of the ISD or IASD form and then demanding the flux to be well quantized, which is a condition that also depends on the complex structure moduli and the complex dilaton \cite{cu}. 

In the present paper we will follow the second approach, imposing the quantization conditions (\ref{quant2}) in order to find solutions for a flux of the form (\ref{G3}). We leave the general derivation and final expression of the flux quantization conditions to Appendix I. Here we just present some simple examples that allow to complete the global $\T^6/\inte_4$ construction. Indeed, let us choose $\tau_1= \tau_2 = \tau_3 = \tau_4 = i$ and impose $A,B$ to be real numbers. Conditions (\ref{quant2}) reduce to
\beq
{A \pm B \over 4\pi^2 \a^\prime} \in N_{\rm min} \inte \quad \Raw \quad
\left\{
\begin{array}{rcl}
\tilde{A}+\tilde{B} & = & 4n \\
\tilde{A}-\tilde{B} & = & 4n'
\end{array}
\right., \quad n, n' \in \inte
\label{cond}
\eeq
where $N_{\rm min}$ is the minimum amount of flux that has to be turned on, in order to satisfy the quantization conditions of fractional cycles on $\inte_4$. In our case $N_{\rm min} = 4$, where we are also taking into account the subtleties mentioned above regarding toroidal orientifolds (See Appendix I for more details). Finally, $\tilde{A} = A/(4\pi^2 \a^\prime)$, $\tilde{B} = B/(4\pi^2 \a^\prime)$, are flux components conveniently normalized.

The D3-brane charge\footnote{In our conventions each D3-brane in the covering space has charge +1. Hence in eq.(\ref{RRcharge-}) $\cam_6$ refers to the covering space for the orientifold, i.e., $\T^6$.} of such $G$-flux will be given by
\beq
Q_{\rm flux} = - 4 \left(\tilde{A}^2 + \tilde{B}^2\right) = - 32 \left( n^2 + n'^2 \right).
\label{D3charge}
\eeq
This negative D3-brane charge will help cancelling the untwisted D3-brane RR tadpole, which is the only RR tadpole left to be cancelled in our global construction. Indeed, the amount of positive D3-brane charge carried by the D3-branes and O3-planes in our model is given by $16 + 4R + 32(b-1)$, and has to be cancelled by the contribution (\ref{D3charge}). This amounts to
\beq
8 \left(n^2 + n'^2\right) - R = 4 + 8 \cdot (b-1),
\label{untwistedflux}
\eeq
which can be satisfied for an infinite numbers of possibilities, some of them being $b=0$, $n=\pm 1, n'=\pm 1, R=4$, or $b=1/2$, $n=\pm 1, n'=0, R=0$, etc.

\subsubsection{Final spectrum}

Let us summarize our results by writing down the final spectrum of our Pati-Salam-like model in table \ref{final}. For definiteness, here we pick the particular choice of discrete B-field $b=1/2$. The visible part of the spectrum will concerns the D3-branes at the origin of $\T^6/\inte_4$, just as in the local construction presented before. The extra D7-branes and D3-branes introduced in order to cancel twisted $k=2$ and untwisted tadpoles will give rise to extra gauge groups and chiral fermions. However, since they are located in a different orientifold fixed point, they will not introduce any extra massless particle charged under the initial Pati-Salam, and thus can be seen as a hidden sector of the theory.

\begin{table}[htb]
\renewcommand{\arraystretch}{1.5}
\begin{center}
\begin{tabular}{|c||c|c|}
\hline
Visible sector & {\bf Gauge group} & {\bf Chiral multiplets} \\
\hline
\hline
$D3-D3$ & $SU(4) \times SU(2) \times SU(2)$ &  
$\begin{array}{c} \vspace*{-.1cm}
2 (4,2,1) + 2 (\bar{4},1,2) \\
(1,2,2) + (10,1,1) + (\bar{10},1,1)
\end{array}$\\
\hline
\hline
Hidden sector & {\bf Gauge group} & {\bf Chiral multiplets} \\
\hline
\hline
$D3'-D3'$ & 
$SU(R) \times SO(R+2)^2$ & 
$\begin{array}{c} \vspace*{-.1cm}
2(R,R+2,1) + 2(\bar{R},1,R+2)\\
(1,R+2,R+2) +  A_R +  \bar{A}_R
\end{array}$\\
\hline
$D7_3-D7_3$ & $SU(4) \times USp(4)^2$ &
$2(4,4,1) + 2(\bar{4},1,4) + (1,4,4) + 10 + \bar{10}$ \\
\hline
$\begin{array}{c} \vspace*{-.1cm}
D3'-D7_3 \\
D7_3-D3'
\end{array}$
& &
$\begin{array}{c} \vspace*{-.1cm}
(R,1,1;1,4,1) + (\bar{R},1,1;1,1,4) \\
(1,R+2,1;4,1,1) + (1,1,R+2;\bar{4},1,1)
\end{array}$\\
\hline
\end{tabular}
\caption{\small Open string spectrum near the origin of the global $\T^6/\inte_4$ model (visible sector), and on the fixed point at $(0,0,(1+i)/2)$ (hidden sector) for the choice of discrete B-field $b=1/2$.
\label{final}}
\end{center}
\end{table}
Notice that the D-brane sector is totally supersymmetric, and so is the spectrum before taking into account the effect of the flux. Indeed, the flux is the only source of SUSY-breaking, in contrast with the models in \cite{ua,aadds} which contained anti-D-branes in order to cancel RR tadpoles. As a result, there are no unbalanced forces between pairs of D-branes, which may induce effective potentials for the D-brane position moduli and extra sources of SUSY breaking in the visible sector.\footnote{In \cite{ua,aadds}, there are anti-branes which are stuck at different fixed points. Therefore, the branes and anti-branes do not annihilate even though there is an attractive force between them. However, this attractive force will generate a potential for the K\"ahler moduli of the orbifold.} Furthermore, the scalar potential generated by the flux for the complex structure moduli fields is at its minimum, since we are turning on IASD components only. Indeed, the only instability of the system comes from the excess of D-brane tension, more specifically $2\times(32 + 4R)$ units of D3-brane tension, which generates an NSNS tadpole typical of non-supersymmetric systems. Notice that this kind of tadpole is the one present in the recent proposals for obtaining de Sitter vacua from string theory \cite{kklt,silver}, before all moduli have been stabilized. As discussed in Appendix II, the precise effect of this tadpole strongly depends on the warp factors of the flux compactification.

\section{Phenomenology}\label{Phenomenology}

In the present section we study some phenomenological features of the flux-induced SUSY breaking scenario. In particular, we focus on the $\T^6/\inte_4$ orientifold model constructed in Section \ref{ModelBuilding}. Our main concern is the effective theory in the visible sector of table \ref{final}, i.e., the D3-brane system located at the origin of the orientifold. Notice that, in the global non-supersymmetric model that we have built, the only explicit source of supersymmetry breaking is given by the IASD $G_3$ flux. Hence, in principle we should not bother about extra sources of SUSY breaking as, e.g., distant anti-D-branes. 

In order to perform the phenomenological analysis, we first briefly review the results in \cite{ciu,ggjl}. In particular, we will focus on the explicit expressions for scalar masses and soft terms derived in \cite{ciu}, and then specialize to the case at hand. The soft-term expressions, as well as the flux quantization conditions, allow us to address a series of phenomenological issues. Some of these issues depend on the local details around the D3-brane location, while some other on the global features of the configuration. As an example of the latter, we consider the distribution of models in the space of well-quantized fluxes that allow for a correct electroweak symmetry breaking pattern.

\subsection{Soft Lagrangian for Flux-Induced Supersymmetry Breaking}

For completeness and to set our notation, here we briefly review the results of \cite{ciu,ggjl} regarding the soft Lagrangian for the flux-induced supersymmetry breaking scenario. In particular, we follow the notation and expressions of \cite{ciu}. We refer the readers to the original papers for the derivation of these results.
%\footnote{The working hypothesis of \cite{ciu} is four-dimensional Poincar\'e invariance, which we will also assume to be the case in the present analysis.} 
In the standard notation of the soft supersymmetry Lagrangian (see, e.g., \cite{bim,susyrev}), the dimensionful couplings on the worldvolume of a D3-brane, and which are induced by the supergravity background around it are given by
\begin{eqnarray}
m_{i \bar{j}}^2\ & =&\ 2K_{i\bar{j}}-\chi_{i\bar{j}}+g_s (\pim
\tau)_{i\bar{j}}  
\nonumber
\\
B_{ij}\ &=& \ 2K_{ij}-\chi_{ij}+g_s(\pim \tau)_{ij} \nonumber
\\
A^{ijk}\ &=& \ -h^{ijk} {{g_s^{1/2}}\over {\sqrt{2}}}\ G_{123}
 \nonumber
\\
C^{ijk}\ &=& +h^{ijl} {{g_s^{1/2}}\over {2\sqrt{2}}}(S_{lk}-(A_{{\bar l}{\bar 
k}})^*)
\nonumber
\\
M^a\ &=& \  {{g_s^{1/2}}\over {\sqrt{2}}}\ G_{123} \nonumber
\\
\mu _{ij} \ & =& \  -{{g_s^{1/2}}\over {2\sqrt{2}}} S_{ij} 
 \nonumber
\\
M_{g}^{ia}\ &=&\ {{g_s^{1/2}}\over {4\sqrt{2}}}
\epsilon_{i\bar{j}\bar{k}} A_{{\bar j}{\bar k}}
\label{softgen}
\end{eqnarray}
where the symmetric and antisymmetric tensors $S_{ij}$ and $A_{\bar{i}\bar{j}}$ are constructed from the 3-form flux $G_3$ as:
\begin{eqnarray}
S_{ij} &=& 
\frac{1}{2}(\epsilon_{ikl}G_{j\bar{k}\bar{l}}+
\epsilon_{jkl}G_{i\bar{k}\bar{l}}) \nonumber\\
A_{\bar{i}\bar{j}} &=&
\frac{1}{2}(\epsilon_{\bar{i}\bar{k}\bar{l}}G_{kl\bar{j}}-
\epsilon_{\bar{j}\bar{k}\bar{l}}G_{kl\bar{i}}) \nonumber
\end{eqnarray}
and likewise for $S_{\bar{i}\bar{j}}$ and $A_{ij}$. The tensors $K$, $\chi$, and $\tau$ come from the power expansion of the metric, dilaton and five-form flux, respectively, around the location of the
D3-branes. As such, they are constrained by the equations of motion of these fields. Combining these constraints, one obtains the following sum-rule for the scalar masses \cite{ciu}:
\begin{eqnarray}
m_1^2 \, +\,  m_2^2\,  +\, m_3^2\, =\, 
 \frac{g_s}{2}\, \left[ \,  \vert G_{123} \vert^2\, +\,
\frac{1}{4}\, \sum_{ij}\, (\, \vert S_{ij} \vert^2\, +\, 
\vert A_{\bar{i}\bar{j}}\vert^2\, )
 \nonumber 
\right. \\  \left. 
  -\,  \preal\, (\, G_{123}G_{\bar{1}\bar{2}\bar{3}}\, +\,
\frac{1}{4}\, S_{lk}S_{\bar{l}\bar{k}}\, +\, 
\frac{1}{4}\, A_{lk}A_{\bar{l}\bar{k}})\, \right]  
\label{genmass}
\end{eqnarray}

Since the matter fields are localized on the brane which preserves ${\cal N}=1$ supersymmetry, their interactions presented above should also be understandable in terms of ${\cal N}=1$, $D=4$ supergravity. From this perspective, the fluxes could be viewed as sources of various auxiliary fields in the supergravity Lagrangian. In particular:

\begin{itemize}

\item 
The $G_{123}$ component of the flux is proportional to the auxiliary component of the dilaton, $F_S$. This gives rise to gaugino masses and trilinear couplings through the operators $\int d \theta^2 S WW$, and $\int d \theta^2 S \phi_i \phi_j \phi_k$, respectively. 

\item
On the other hand, the flux $A_{\bar{k} \bar{l}}$ is part of the D term of a vector super-multiplet. The coupling $M^{ia}_g$ has its origin in the operator $\int d \theta^2 f(S, \phi) W W$, where $f$ is the gauge kinetic function. From this term we obtain $M^{ia}_g \propto \partial_i f (S, \phi) D$. Comparing this with the expression of $M^{ia}_g$ in terms of flux, we can identify $D \propto \epsilon_{i\bar{j}\bar{k}} A_{{\bar j}{\bar k}}$. There is also a term in the scalar potential $1/2 (D f(S, \phi) D)$, where $D \propto \phi_i^* t_{ij} \phi_j$. Expanding the gauge kinetic function and using the identification we have just made, we obtain the term proportional to $A_{{\bar j}{\bar k}}^*$ in $C^{ijk}$. 

\item
Turning on flux $S_{ij}$ does not break ${\cal N}=1$ supersymmetry on the D3 brane. This is evident at the Lagrangian level. The effect of this flux is to generate a $\mu$-term in the superpotential. The $|F_i|^2$ term derived from the superpotential is then responsible for the term proportional to $S_{ij}$ in the coupling $C^{ijk}$. Hence, $C^{ijk}$  contains both a supersymmetry preserving part and a supersymmetry breaking part (proportional to the flux component $A_{{\bar j}{\bar k}}^*$ ). 

\item
Both $M^{ia}_g$ and the supersymmetry breaking part of $C^{ijk}$ are somewhat atypical supersymmetry breaking terms. On the one hand, the terms $C^{ijk}$ are considered to be soft only if there is no chiral multiplet which is a singlet of the gauge group. This is a condition that the Minimal Supersymmetric Standard Model (MSSM) actually satisfies. However, those terms are usually not included as part of the soft supersymmetry breaking Lagrangian of the MSSM. On the other hand, gauge symmetry requires the existence of adjoint light matter fields for a non-vanishing $M^{ia}_g$. Therefore, including $M^{ia}_g$ in the soft lagrangian generically requires one to go beyond the matter content of the MSSM and postulate the existence of some adjoint matter fields. In the case of flux-induced couplings, we see that the supersymmetry breaking part of $C^{ijk}$ and $M^{ia}_g$ are proportional to the same flux $A_{\bar{j} \bar{k}}$. Therefore, the SUSY-breaking part of $C^{ijk}$ would also vanish when we have only the MSSM matter content. 

\end{itemize}

Let us now focus on the soft terms present on a D3-brane placed on top of a $\cpx^3/\inte_4$ singularity. The non-zero components of the $G_3$ flux allowed by the orbifold symmetries are $G_{123}$ and $G_{\bar{1}\bar{2}3}$:
\begin{equation}
G_3 = G_{123}~dz^1 \wedge dz^2 \wedge dz^3
+ G_{\bar{1}\bar{2}3} ~d \bar{z}^1 \wedge d \bar{z}^2 \wedge dz^3 
\end{equation}
Therefore, $A_{ij}=A_{\bar{i}\bar{j}}=S_{\bar{i}\bar{j}}=0$ and $S_{ij}=0$, except $S_{33} = 2 G_{\bar{1}\bar{2}3}$.
As a result, the scalar masses are
\begin{equation}
m_1^2 + m_2^2 + m_3^2 = \frac{g_s}{2} \left[ \, |G_{123}|^2 +
|G_{\bar{1}\bar{2}3}|^2 \, 
\right]
\label{scalarmass} 
\end{equation}

It is important to distinguish between scalar masses and soft masses. In eq.(\ref{scalarmass}), there are two contributions to the scalar masses. One of them is proportional to $|G_{\bar{1}\bar{2}3}|^2 \propto |\mu|^2$, which is the supersymmetric mass term. The soft scalar masses contribution $m_0$, which breaks supersymmetry and does not arise from the superpotential, corresponds to the piece proportional to $|G_{123}|^2$. 

The other soft terms are:
\begin{eqnarray}\label{soft}
A^{ijk}\ &=& \ -h^{ijk} {{g_s^{1/2}}\over {\sqrt{2}}}\ G_{123}
 \nonumber
\\
M^a\ &=& \  {{g_s^{1/2}}\over {\sqrt{2}}}\ G_{123}.
\label{softterms}
\end{eqnarray}
Finally, we also have a non-vanishing $\mu$-term:
\begin{eqnarray}
\mu _{ij} \ & =& \  -{{g_s^{1/2}}\over {\sqrt{2}}} G_{\bar{1}\bar{2}3}
\delta_{i3} \delta_{j3} . 
%M_{g}^{ia}\ &=& 0
\label{muterm}
\end{eqnarray}

Notice that the spectrum presented in eqs.(\ref{scalarmass}) and (\ref{softterms}), assuming universal soft masses, is identical to the spectrum of dilaton dominated SUSY breaking scenario with vanishing cosmological constant \cite{dilatondomination}. The soft parameters in this scenario obey the simple relations $A^{ijk} = - h^{ijk} M^a$ and $|M^a| = \sqrt{3} m_0$. This is not a surprise at all since the supersymmetry breaking $G_{123}$ flux can be understood as the source of the dilaton $F$-term $F_S$. In this sense, the soft term pattern of our string construction, as well as its associated low energy physics, can be handled by means of a $D=4$ Effective Lagrangian approach. Notice, however, that there is a remarkable fact of the flux-induced SUSY breaking scenario. Namely, we have an {\em explicit source of supersymmetry breaking} (in this case $F_S$) which is {\em determined by the background flux instead of being a phenomenological input}. In the following we will exploit this fact, with the aim of learning some lessons from the flux-induced SUSY-breaking scenario. As we will see, although we perform our analysis on the particular $\T^6/\inte_4$ model constructed in the previous section, the final conclusions are rather model-independent.

\subsection{Electroweak Symmetry Breaking}

An important feature of the model constructed in Section \ref{ModelBuilding} is that the background flux $G_3$ provides a source for {\it both} the soft parameters and the $\mu$-term. As discussed below, these essential quantities in electroweak symmetry breaking (EWSB) generically need to be fine-tuned, at least at the level of one percent, in order to achieve a phenomenologically viable electroweak symmetry breaking. It is thus interesting to see if the fact that these parameters have the same microscopic origin (i.e., the flux) can give us some insight into this fine-tuning problem involved in EWSB.

Our approach is different in spirit from the conventional studies of EWSB in a SUSY-breaking scenario. Typically, a particular SUSY-breaking scenario only provides a mechanism of generating soft 
SUSY-breaking terms. The $\mu$-term is then treated as a free parameter, which is then fixed by requiring a correct electroweak symmetry breaking at low energies. Now, when such SUSY-breaking scenario is described by an actual microscopic theory, one may wonder if there are some constraints on the otherwise `free parameters', such as the $\mu$-term. In the particular case of flux-induced SUSY breaking this turns out to be the case.\footnote{Similar relations between the $\mu$-term and the soft parameters also appear in the context of Scherk-Schwarz supersymmetry breaking in 5D orbifold models \cite{Kiwoon}. We thank Kiwoon Choi for pointing these references to us.} For instance, as explained in Section \ref{ModelBuilding} the components of the flux $G_3$ must be properly quantized in terms of the internal geometry of $\cam_6$. This implies that, although $\mu$ and $M^a$ are given by different components of the flux, in an actual model they are related and constrained by the flux quantization conditions.

In the model at hand, a 3-form flux $G_3$ has to satisfy eight different quantization conditions, which are presented in Appendix I. Notice that, in general, $G_3$ is complex-valued, and hence it will give rise to complex soft term in our model. In an scenario where the superpartner masses are of the order of TeV scale, however, low energy measurements on CP violating observables such as the electric dipole moment \cite{EDM} generically constrain the supersymmetry breaking parameters to be real. In order to satisfy this phenomenological constraint, we will consider well-quantized fluxes $G_3$ which 
have only real components. This may seem to be a rather strong assumption in the space of fluxes. However, in our analysis we will be mainly interested in how the conditions required for electroweak symmetry breaking are realized in this scenario. As will become obvious later, the mechanism we discuss will only depend on the fact that the fluxes are quantized, and not on them being real. Hence, it would also work in a more general scenario.\footnote{As, e.g.,  the one recently suggested in \cite{fluxed}.} 

With the above assumptions, the quantization conditions then reduce to eq.(\ref{cond2}) in Appendix I, which in turn imply
\be
|G_{123}|=\left( 1+\frac{1}{g_s} \frac{m}{n} \right) M_{soft}, \quad \quad
|G_{\bar{1} \bar{2}3}| = \left( 1-\frac{1}{g_s} \frac{m}{n} \right) M_{soft},
\ee
for some integers $n$ and $m$. Here we have made use of the fact that $\pim \tau_4 = 1/ g_s$, and we have factored out the scale $M_{soft} = 8 n \pi^2 \alpha'/\sqrt{Vol(\T^6/\inte_4)}$. Notice that the values that the fluxes can take are discretized, since $g_s$ is also fixed by the flux.

Electroweak symmetry breaking will be induced, as usual, by the radiative corrections to the soft Higgs masses. There are two low energy outputs of the electroweak symmetry breaking. They are $v^2=v_u^2+v_d^2$, or equivalently $m_Z^2$, and $\tan \beta = v_u/v_d$. The only value that is currently measured is $m_Z^2$. The value of $\tan \beta \sim (m_{H_u}^2 + m_{H_d}^2+2 |\mu|^2)/ \mu B$ is determined by $\mu B$ which in our model depends on $K_{ij}$, $\tau_{ij}$ and $\chi_{ij}$, and so it is not constrained by the local equations of motion derived in ref\cite{ciu}. Therefore, we will take $\tan \beta$ as a free parameter at this stage.\footnote{In a {\it global} model such as the one presented here, generically there will be constraints on $\mu B$ from the equations of motion. Although $\tan \beta$ is an important parameter for various aspects of supersymmetry phenomenology \cite{susyrev}, its precise value does not affect the tuning involved in electroweak symmetry breaking significantly. Therefore, for our purpose it is good enough to assume that $\tan \beta$ is a free parameter. It would be interesting to study the constraints on $\tan \beta$ from the equations of motion in some explicit global models.}
 
Roughly speaking, the condition for a correct electroweak symmetry breaking is that it reproduces the measured Z-boson mass. As we commented above, since we have a prediction for $\mu$, the condition becomes non-trivial. We will devote the rest of this section studying how this condition can be satisfied. Including the effect of RGE running, the requirement can be written as \cite{ewsb}:
\be
m_Z^2 \simeq  C_{\mu}|\mu|^2 + C_{1/2} m_{1/2}^2 + C_0 m_0^2 + C_A 
A_t^2 + C_{A \lambda} A_t m_{1/2}.
\ee
where $m_{1/2}$ and $A_t$ are the universal gaugino mass and trilinear coupling respectively. The numerical coefficients $C_i$'s arise from integrating the renormalization equations for the soft parameters and $\mu$ and minimizing the Higgs potential. As a result, they depend on several other parameters, such as top Yukawa coupling (or $m_{top}$), $\tan \beta$, and $\Lambda_{\mbox{\footnotesize{UV}}}$ which is the scale at which supersymmetry is broken. In our setup, $\Lambda_{\mbox{\footnotesize{string}}} \sim \Lambda_{\mbox{\footnotesize{UV}}} \sim 10^{11} $ GeV \cite{ciu}. For the sake of concreteness, let us pick the MSSM values which are deduced from this $\Lambda_{\mbox{\footnotesize{UV}}}$ and $\tan \beta = 5$. The coefficients then read \cite{ewsb}: 
\be
C_{\mu} \simeq -1.7, \ \ C_{1/2} \simeq 2.6,, \ \ C_0 \simeq 0.3, \ \
C_A \simeq 0.2, \ \ C_{A \lambda} \simeq -0.6. 
\ee

The general statement about the required fine-tuning for electroweak symmetry breaking is obvious here. Indeed, we could characterize the size of the soft parameters by some scale $M_{soft}$. On the one hand, non-observation of superpartners at LEP-II puts various lower bounds on $M_{soft}$ already. Moreover, in the MSSM, a heavier Higgs mass which satisfies the LEP lower bound \cite{lephiggs}  would almost necessarily require at least some of the soft parameters to be several times larger than $M_Z$ \cite{Carena:2002es}. On the other hand, the level of cancellation which is required to achieve electroweak symmetry breaking increases quadratically with $M_{soft}$, which generically requires a $1 \%$ cancellation. More precisely we could use, as a measure of the level of fine-tuning, the variation of the Z-boson mass resulting from a change in the soft masses and the $\mu$-term:
\be
\frac{\delta m_Z^2}{m_Z^2} = \frac{m_i^2}{m_Z^2} C_i \left(\frac{\delta m_i^2}{ m_i^2}\right).
\label{quadratic-finetuning}
\ee
By requiring $\delta m_Z^2 / m_Z^2 \leq 1$, we obtain $\delta m_i^2 / m_i^2 \sim m_Z^2 /(C_i m_i^2) \sim 1 \%$ if $m_i \sim$ TeV. This shows the $1 \%$ tuning of the electroweak symmetry breaking in the MSSM, as well as the quadratic grow of this fine-tuning. 

Taking these facts seriously, there is a somewhat significant tension between electroweak symmetry breaking and the experimental Higgs mass bound within the framework of the MSSM. However, it is important to keep in mind the assumptions underlying such an argument. First, the above reasoning is based on the tuning of the soft parameters of  an {\em effective} theory -- the soft SUSY Lagrangian. Given a {\it microscopic} theory of supersymmetry breaking, the soft terms are determined by fundamental quantities of this underlying theory. A priori, this effective fine-tuning could look different when expressed in terms of these fundamental parameters. In this case, the quantity $\delta m_i^2 / m_i^2$ may not accurately characterize the likelihood of finding a correct EWSB solution. Second, we are supposing that the $\mu$-term, which is an important part of the Higgs mass, is a free parameter that can be tuned.

For instance, the above assumptions may not apply when the soft parameters and the $\mu$-term are quantized. In fact, one of the main results of this section is to show that at least in the class of string models with flux induced supersymmetry breaking that we studied, the breaking of electroweak symmetry does not necessarily suffer from a quadratic fine-tuning. In some cases, the degree of fine-tuning may even decrease with a higher scale of $M_{soft}$.

Let us be more explicit. Using Eqs.~(\ref{scalarmass}), (\ref{soft}) and (\ref{muterm}), we have for this model
\be
m_Z^2 = (-1.7 \lambda^2 + 3.5) M_{soft}^2
\label{ewsbcond}
\ee
where $\lambda$ is the ratio of two components of the IASD flux
\be
\frac{G_{\bar{1} \bar{2} 3}}{ G_{123}} \equiv \lambda=\frac{1-g_s
n/m}{1+g_s  n/m},
\label{famous}
\ee
Increasing soft masses $\propto M_{soft}^2 \gg m_Z^2 $ would require us to adjust $(-1.7 \lambda^2 + 3.5)\sim 0$ in order to achieve electroweak symmetry breaking. Therefore, $n/m <0$. It also means that for $M_{soft}$ large, $\lambda$ is almost fixed. Since we have the freedom of adjusting $\lambda$ at will by adjusting the ratio $n/m$, we expect to find models satisfying the requirement of Eq.~(\ref{ewsbcond}) for various values of $M_{soft}$. 

We would like to assess further the level of difficulty in obtaining string vacua leading to phenomenologically viable EWSB as we vary $M_{soft}$. To be precise, we measure the degree of fine-tuning by the number of string vacua in which Eq~(\ref{ewsbcond}) is satisfied.\footnote{This stringy naturalness point of view on the fine-tuning problem is similar in spirit to \cite{Bousso:2000xa,Susskind:2004uv,Douglas:2004qg} on the scale of supersymmetry breaking in the string theory landscape. For some other applications of the stringy naturalness principle, see \cite{Dine:2004is,Silverstein:2004sh,Conlon:2004ds,Kumar:2004pv,dgkt}.}
In particular, we are interested in  the distribution of solutions as a function of $M_{soft}$. These results can then be compared with the standard expectation of a quadratic fine-tuning, i.e., the number of solutions going as $M_{soft}^{-2}$.

\begin{figure}[ht]
\centering
\includegraphics[scale=0.575,angle=270]{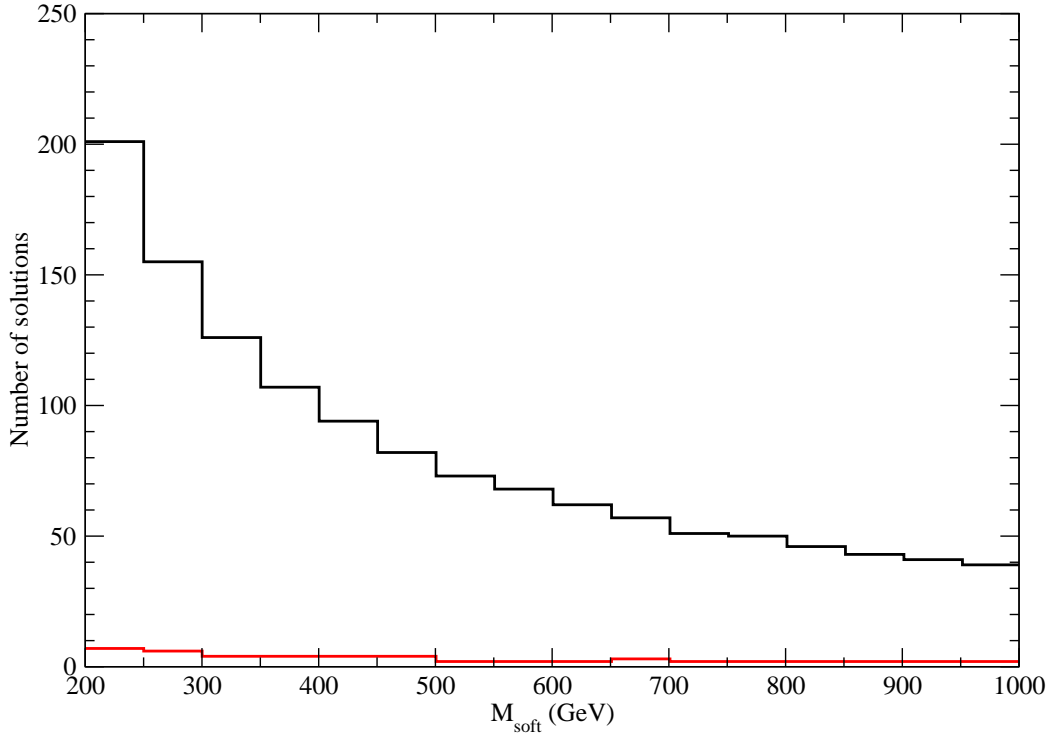}
\caption{Number of solutions as a function of $M_{soft}$ if $\Lambda_S$ is
independent of $n$.
The black (upper) line corresponds to $\Lambda_S=10$ GeV. The red (lower) line
corresponds to $\Lambda_S = 50$ GeV.
  \label{ms}}
\end{figure}

\begin{figure}[h!]
\centering
\includegraphics[scale=0.575,angle=270]{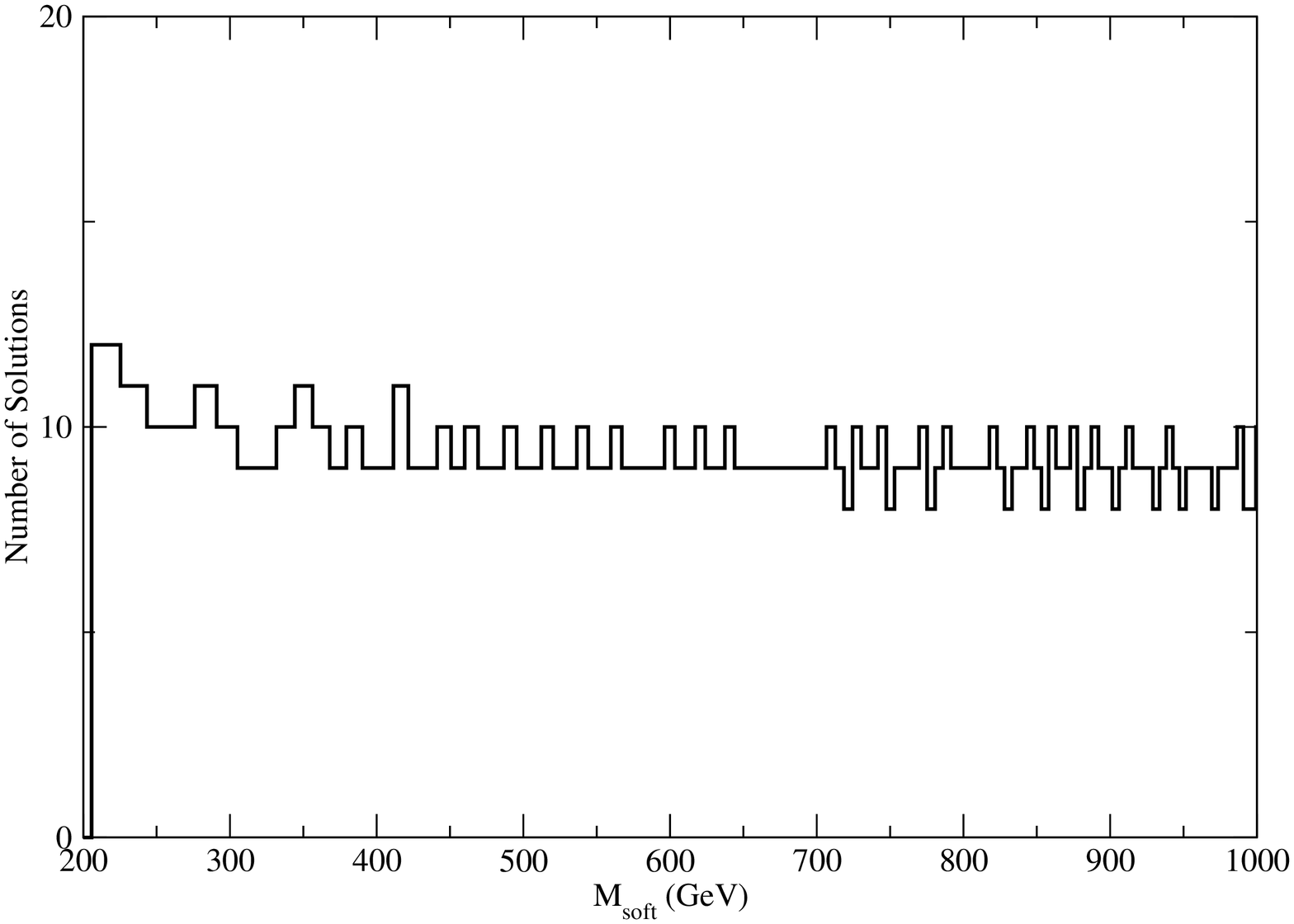}
\caption{Number of solutions as a function of $M_{soft}$ for $\alpha=2$, assuming $m_X = m_Z$. \label{msn2}}
\end{figure}

\begin{figure}[h!]
\centering
\includegraphics[scale=0.575,angle=270]{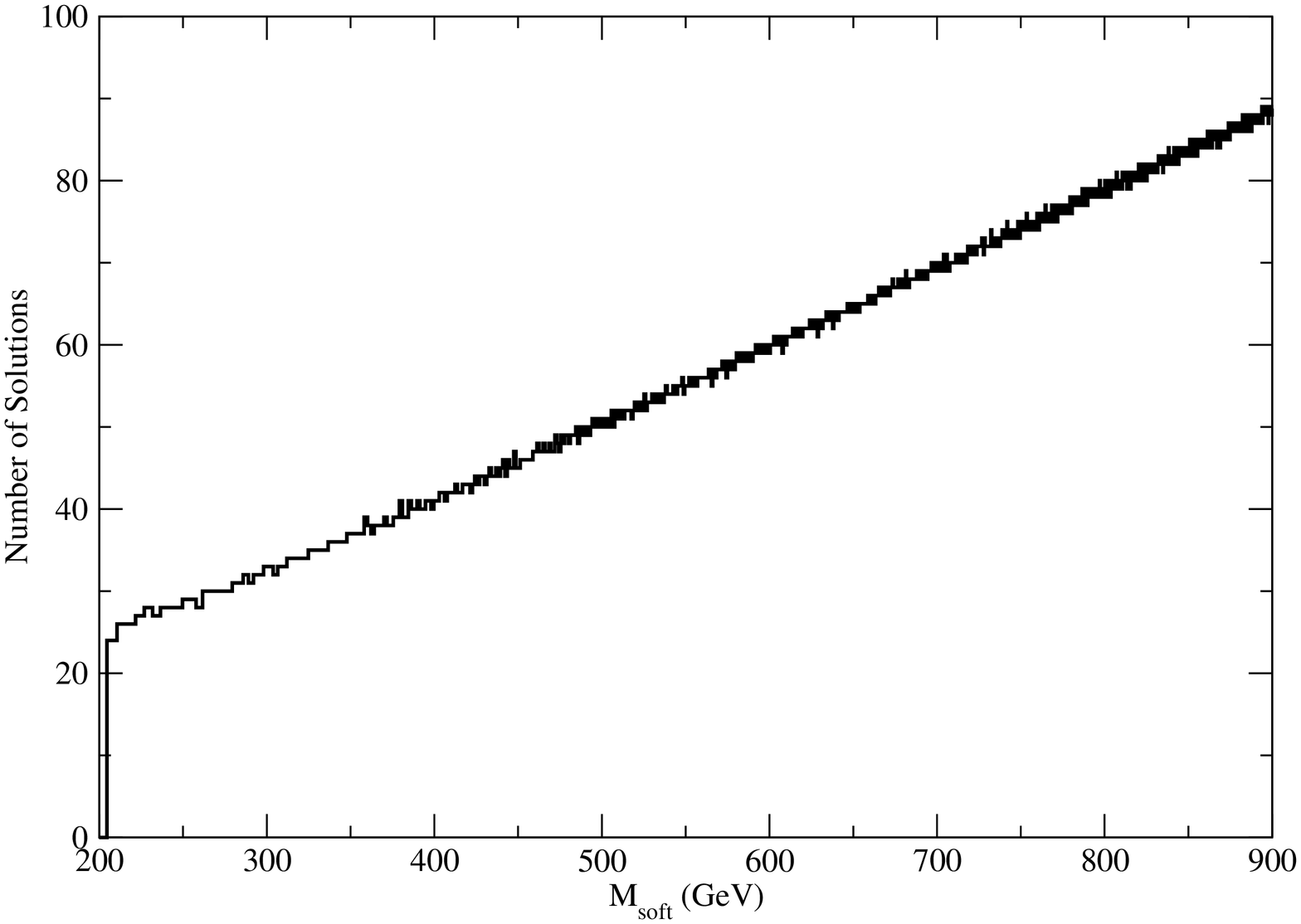}
\caption{Number of solutions as a function of $M_{soft}$ for $\alpha=3$, assuming $m_X = m_Z$. \label{msn3}}
\end{figure}

Let us count the number of vacua with flux quanta $n$ and $m$ which satisfy approximately\footnote{To qualify as an approximate solution, the difference between the experimental value of the Z-boson mass and the value calculated from Eq.~(\ref{ewsbcond}), denoted as $\delta m_Z$, should be sufficiently smaller than $m_Z$.} Eq.~(\ref{ewsbcond}). It is clear from the definition that $\lambda$ depends on $(n,m)$. The functional dependence of $M_{soft}$ on the flux quanta is slightly more subtle. First, note that the value of $M_{soft}$ depends on the 3-form flux density, which in turn depends on the flux quantum $n$ and the volume of the corresponding 3-cycle. For our model at hand,
\begin{equation}
M_{soft} = 8 n \pi^2 \alpha'/\sqrt{Vol(\T^6/\inte_4)} \equiv n \Lambda_S.
\end{equation}
In addition to the explicit $n$ dependence, $\Lambda_S$ will be in general a function of $n$. The precise functional form of $\Lambda_S$ depends on the mechanism which stabilizes the K\"ahler moduli. For instance, in the KKLT \cite{kklt} scenario, the runaway behavior of the overall K\"ahler modulus due to uncanceled NSNS tadpoles is counteracted by non-perturbative effects, thus stabilizing the moduli at a finite value. Increasing the value of $n$ would increase the NSNS tadpoles and so we expect the stabilized value of ${\rm Vol}(\T^6/\inte_4)$  to increase with $n$. In this case, $\Lambda$ should be a decreasing function of $n$. More generally, let us parametrize the unknown mechanism that stabilizes the K\"ahler moduli by writing
\begin{equation}
\Lambda_S \equiv n^{1/\alpha-1} m_X ~.
\end{equation}
In other words,
\begin{equation}
M_{soft} = n \Lambda_S = n^{1/\alpha} m_X
\end{equation}
where $m_X$ is a $n$-independent mass scale. The tuning involved in EWSB is now
\beq
\frac{\delta m_Z^2}{m_Z^2}\ =\ 
\frac{\partial }{\partial n} \log \left[(-1.7 \lambda^2 +
3.5)\, n^{2/\a}\, m_X^2\right]\cdot \delta n 
\ \propto\ 
\left( n^{2/\a} \frac{m_X^2}{m_Z^2} + \frac{2}{\a} \right) \frac{\delta n}{n},
\label{stringy-finetuning}
\eeq
Therefore, the degree of fine-tuning depends on the value of $\alpha$ (hence the precise mechanism which stabilizes K\"ahler moduli) and is not necessarily quadratic in $M_{soft}$. More precisely, the leading behaviour of this fine-tuning on $M_{soft}$ will be given by $M_{soft}^{2-\a}$. For illustrative purpose, let us consider the following cases:
\begin{itemize}

\item $\alpha=1$: In this case, the volume modulus is stabilized at a value independent of the flux quanta $n$. Although the number of flux vacua leading to a viable EWSB decreases with $M_{soft}$ as depicted in Figure \ref{ms}, the tuning involved is {\it linear} rather than quadratic in $M_{soft}$.

\item $\alpha=2$: In this case, the number of flux vacua satisfying the EWSB condition of Eq.~(\ref{ewsbcond}) stays approximately {\it constant}, as shown in Fig.~\ref{msn2}. Again, this is different from the quadratic fine-tuning one would expect based on effective field theory arguments.

\item $\alpha>2$: In this case, the number of solutions actually {\it increases} with increasing $M_{soft}$. This means that it is {\it easier} to find vacua leading to a correct EWSB with a higher scale of $M_{soft}$, which is exactly opposite to the conclusion one would have made based on the conventional intuition of fine-tuning. Fig.~\ref{msn3} shows the likelihood of correct EWSB for $\alpha=3$.

\end{itemize}

Notice that the quadratic behaviour on $M_{soft}$ is only recovered when $\a \raw 0$. Quite amusingly, this corresponds to the case where ${\rm Vol}(\T^6/\inte_4) \raw \infty$, that is, where there is no stabilization mechanism for the K\"ahler moduli that prevents the theory to be driven to the decompactification limit.

%%%%%%%%%%%%%%%%%%%%%%%%%%%%%%%%%%%%%%%%%
%%%%%%%%%%%%%%%%%%%%%%%%%%%%%%%%%%%%%%%%%

%%%%%%%%%%%%%%%%%%%%%%%%%%%%%%%%%%%%%%%%%%%%%%%%%
%%%%%%%%%%%%%%%%%%%%%%%%%%%%%%%%%%%%%%%%%%%%%%%%

%%%%%%%%%%%%%%%%%%%%%%%%%%%%%%%%%%%%%%%%%%%%%%%%%
%%%%%%%%%%%%%%%%%%%%%%%%%%%%%%%%%%%%%%%%%%%%%%%%

It is also interesting to relate our results to the recent proposals of statistically favored high scale supersymmetry breaking \cite{Susskind:2004uv,Douglas:2004qg}. Suppose we have a set of supersymmetry breaking order parameters $X_i$, $i=1,\dots,n_X$, after putting in the constraints of electroweak symmetry breaking, the volume of parameter space is \cite{Douglas:2004qg}
\be
d \mu \propto M_{soft} d(M_{soft}) \int d^{n_X} X_i \delta(M_{soft}^2 -
\sum X_i^2) \frac{\Lambda_{EW}^2}{M_{soft}^2} \rightarrow \Lambda_{EW}^2
M_{soft}^{n_X-3} d(M_{soft})
\ee
where the $\delta$-function impose the condition of all
supersymmetry breaking scale. The number of states satisfy these
constraints will be $n_{sol} \sim d \mu \times \rho$, where $\rho$ is
the density of states. For example, if $\rho = $constant, there are
more solutions at higher scales if $n_X \geq 3$. One simple
realization of this scenario involves two complex $F$ fields as
the supersymmetry breaking order parameter. 

In our case,  we also have two auxiliary $F$ fields. However, there
are no large splittings between scalar and gaugino masses in this
scenario due to the universal nature of the dilaton coupling. Therefore,
unlike the scenario proposed in ref.~\cite{splitsusy},
compatibility with gauge unification will favor a relatively low
supersymmetry breaking scale $\sim$ TeV.  In this case,
there are strong constraints on the CP violations from, e.g.,
EDM observables \cite{EDM} . Generically, the phases of the supersymmetry breaking
parameters are constrained to be real. Therefore, effectively,
$n_X=2$ in this scenario. On the other hand, as commented earlier in
this section, the distribution of states is not necessarily
uniform but can be more densely populated at relatively higher scales. Therefore, the
number of solutions 
would not necessarily decrease, and may even increase, with 
the scale of 
$M_{soft}$.

It should be emphasized that this is by no means a scan of the string theory landscape (since we
are only exploring a tiny subspace of string vacua) nor do we attach any statistical interpretation to our results (which require assumptions about the precise way in which all moduli are stabilized, and
the measure on the landscape). 
Nevertheless, we provide an explicit example to illustrate that the issue of fine-tuning in electroweak symmetry breaking may be viewed differently when the distribution of consistent string vacua are taken into account. In particular, larger soft masses, at least those which is directly relevant to electroweak symmetry, are {\it not} necessarily disfavored \cite{splitsusy,Susskind:2004uv,Douglas:2004qg}. Moreover, as shown in our analysis, this effect is only relying on the fact that fluxes are quantized (hence $n$ and $m$ take on integer values) and is not sensitive to the explicit form of $\lambda$ as a function of these parameters. Therefore, we expect this 
{\it suppressed} fine-tuning effect to be generic in a large class of models with flux-induced electroweak symmetry breaking.\footnote{At the same time, we have not taken into account effects such as the next-leading order corrections, threshold corrections, high energy input scale of RGE running, etc., on the Higgs potential. Changing $\tan \beta$ will also change the specific numbers used in the condition of electroweak symmetry breaking. We do not expect, however, the suppressed fine-tuning effect to depend on the particular details of the Higgs potential.}

\section{Conclusions}\label{Conclusion}

In this paper, we have studied supersymmetry breaking effects induced on D3-branes at singularities by the presence of NSNS and RR 3-form fluxes. We have constructed some global models of flux compactifications as examples to illustrate the issues of model building and phenomenology involved in this scenario. Although the models presented are not fully realistic, since they do not contain enough number of chiral families to account for the SM flavor structure, our analysis serves as a template for future phenomenological studies of this scenario. In any case, since the soft parameters induced by the fluxes on D3-branes at singularities are universal and independent of the number of families, we expect our conclusions on phenomenological features which are insensitive to them, such as the fine-tuning involved in EWSB, to persist in more realistic models.

In fact, the key feature of this SUSY breaking scenario that allows for a different behaviour from the usual EWSB fine-tuning is the fact that the soft parameters are quantized. In our setup things are particularly simple, since both $\mu$ and soft terms come from the same source, namely the fluxes. In general, there could be other sources for them.  However, we expect our arguments to apply as long as these parameters take on quantized values.

Other than the fine-tuning issue of electroweak symmetry breaking, we have not discussed in depth other phenomenological features of this SUSY breaking scenario. The reason is that many phenomenological issues surrounding supersymmetry breaking, such as flavor and CP observables, involve the family structure of the Standard Model which is not realized in the models at hand. Clearly, it would be interesting to carry out a detailed study of CP observables for flux vacua with realistic flavor structure. However, without a detailed knowledge of the flavor structure, we could only anticipate that this type of models would produce signatures similar to a subclass of mSUGRA scenarios \cite{lhcsusy}. Notice that the spectrum of this class of models is identical to the dilaton dominated supersymmetry breaking scenario. One of the generic features of this class of models is the soft masses and trilinear couplings are universal.\footnote{On the same footing, gaugino masses are also universal because the entire Standard Model is embedded on a single stack of D-branes. This is in general not the case in the magnetized D-brane setup as, e.g., the models in \cite{ms}. For some explicit computations which illustrate this fact see \cite{lrsMSSM}.} As a result, we expect the Bino to be the lightest superpartner. In addition, a naive estimate of the string scale in this class of models gives $M_s \sim 10^{11}$ GeV, and so string scale stable states (such as D-matter \cite{Shiu:2003ta}) could also be candidates for Wimpzillas dark matter \cite{wimpzillas}. Finally, there is no large splitting between gaugino and sfermion masses, which is very different from the split supersymmetry scenario recently proposed in \cite{splitsusy}.  

While this paper was written, ref. \cite{ciu2} appeared where a $\inte_4$ orbifold model involving D3 and D7-branes was constructed, and the flux-induced soft terms induced in such construction were computed. Although there are some similarities with the ${\bf Z}_4$ model presented here, the constructions are not identical. It would be interesting to find a global embedding of \cite{ciu2} and compare the resulting phenomenology with the global models discussed here.

Finally, in the magnetized D-brane setup, the Standard Model does not have to be localized completely at a singularity and so more realistic models can be constructed \cite{ms}. Some of the aforementioned challenges in constructing realistic flux vacua that appeared in models including only D3-branes at singularities are a priori absent, and this seems a promising direction for constructing realistic models with fluxes. It would be interesting to apply the recent results of flux induced SUSY breaking terms for the D3/D7-system computed in \cite{lrs,ciu2} to the three-generation MSSM flux vacua in \cite{ms}.

\bigskip

\bigskip

\bigskip

\noindent
{\bf Acknowledgments:}
We would like to thank Nima Arkani-Hamed, Vijay Balasubramanian, Alex Buchel, Pablo G. C\'amara, Kiwoon Choi, Tony Gherghetta, Luis Ib\'a\~nez, Gordy Kane, Hans-Peter Nilles, and specially Lisa Everett and Angel Uranga for useful comments and discussions.
This work was supported in part by a DOE grant No. DE-FG-02-95ER40896. FM and GS were also supported by NSF CAREER Award No.~PHY-0348093, and a Research Innovation Award from Research Corporation. LW was partially supported by the Wisconsin Alumni Research Foundation. 
We would also like to thank the Aspen Center for Physics for hospitality.

\newpage

\section{Appendix I: $\T^6/\inte_4$ flux quantization conditions}

In this appendix we collect the flux quantization conditions for the particular case of the toroidal orientifold $\T^6/\inte_4$. In a general geometry, these quantization conditions are given by (\ref{quant2}). In the particular case that $\cam_6$ is an orbifold, one needs to integrate the 3-form fluxes $F_3$ and $H_3$ over an integral basis of 3-cycles. Now, an integral basis of 3-cycles for $\T^{2n}/\inte_N$ or $\T^{2n}/\inte_N \times \inte_M$ involves `fractional' cycles, i.e., those which are closed in the orbifold but not in the covering space $\T^{2n}$. If we are not introducing any 3-form flux in the collapsed cycles of $\T^{2n}/\inte_N$, however, the flux quantization conditions can be recasted in terms of the homology of $\T^{2n}$. More precisely, we have
\beq
{1 \over 2 \pi \a^\prime} \int_{\Sigma} F_3 \in 2\pi N_{\rm min} \inte, \quad \quad
{1 \over 2 \pi \a^\prime} \int_{\Sigma} H_3 \in 2\pi N_{\rm min} \inte
\label{quantT}
\eeq
where $\Sig$ is now a 3-cycle of $\T^{2n}$, and $N_{\rm min}$ is an integer which depends on the particular orbifold. In a $\inte_4$ orbifold this number can be seen to be $N_{\rm min}^{\inte_4} = 2$.\footnote{Roughly speaking, $N_{\rm min}$ can be computed from dividing the volume of a 3-cycle on $\T^{2n}$ by its corresponding fractional cycle on $\T^{2n}/\inte_N$. The integral homology of $\T^6/\inte_4$ can be found in, e.g., \cite{blumZ4}.} Considering an {\em orientifold} of the previous construction implies that we have an extra $\inte_2$ action acting on $\T^6$, and hence the previous number $N_{\rm min}$ may be multiplied by 2. This turns out to be the case in the $\T^6/\inte_4$ model discussed in the main text,\footnote{The precise statement goes as follows \cite{fp}. We may either have that the flux quanta along the cycle $\Sig$ are of the form $(2n + 1) N_{\rm min}^{\rm orbi}$ or $2n N_{\rm min}^{\rm orbi}$ ($n \in \inte$) depending on the fact that $\Sig$ passes over an odd or even number of $O3^{(+,+)}$, respectively. In the $\inte_4$ orbifold example considered in the main text we are in the second situation, so we can effectively take $N_{\rm min}^{\rm ori} = 2 N_{\rm min}^{\rm orbi}$. See also the comments in \cite{blum} regarding this problem.} so that at the end of the day we have to impose conditions (\ref{quantT}) with $\Sig$ being any 3-cycle of $\T^6$ and $N_{\rm min} = 4$.

These conditions can be written in a more specific way by using \cite{cu}
\beq
F_3 = - {\pim (\bar{\tau}_4 G_3) \over \pim \tau_4}, \quad \quad 
H_3 = - {\pim G_3 \over \pim \tau_4},
\label{FH}
\eeq
and
\beq
\int_{[a_i]} dz_j = \d_{ij}, \quad \quad \int_{[b_i]} dz_j = \tau_i \d_{ij},
\label{1cycles}
\eeq
where $i,j = 1,2,3$ and $[a_i]$, $[b_i]$ is the basis of 1-cycles on the $i^{th}$ $\T^2$. By using the explicit expression 
\beq
G_3\ =\ A \cdot dz_1 \wedge dz_2 \wedge dz_3\ +\ B \cdot d\bar{z}_1 \wedge d\bar{z}_2 \wedge dz_3
\label{G3ap}
\eeq
for the 3-form flux, conditions (\ref{quantT}) are finally translated into
\beq
\begin{array}{lcl} \vspace*{.3cm}
{\pim (\tilde{A}+\tilde{B}) \over \pim \tau_4} \in 4 \inte & & {\pim [(\tilde{A}+\tilde{B})\bar{\tau}_4] \over \pim \tau_4} \in 4 \inte \\ \vspace*{.3cm}
{\pim [(\tilde{A}+\tilde{B})\tau_3] \over \pim \tau_4} \in 4 \inte & & {\pim [(\tilde{A}+\tilde{B})\tau_3 \bar{\tau}_4] \over \pim \tau_4} \in 4 \inte \\ \vspace*{.3cm}
{\preal (\tilde{A}-\tilde{B}) \over \pim \tau_4} \in 4 \inte & & {\preal [(\tilde{A}-\tilde{B})\bar{\tau}_4] \over \pim \tau_4} \in 4 \inte \\
{\preal [(\tilde{A}-\tilde{B})\tau_3] \over \pim \tau_4} \in 4 \inte & & {\preal [(\tilde{A}-\tilde{B})\tau_3 \bar{\tau}_4] \over \pim \tau_4} \in 4 \inte
\end{array}
\label{condiG3}
\eeq
where we have conveniently normalized the flux components as $\tilde{A} = A/(4\pi^2 \a^\prime)$, $\tilde{B} = B/(4\pi^2 \a^\prime)$. Notice that the complex structure of the first two-tori of $\T^6/\inte_4$ do not enter into these conditions, since the orbifold geometry fixes them to be $\tau_1 = \tau_2 = i$.

These previous equations are simplified by choosing some subspace of flux quanta. For instance, as discussed in the main text, a phenomenologically interesting subspace is given by taking $\tilde{A}, \tilde{B}$ to be real numbers.  The quantization conditions (\ref{cond}) then read

\beq
\begin{array}{rcl}\vspace*{.2cm}
\tilde{A}+\tilde{B} & = & 4\, n_1 \\ \vspace*{.2cm}
(\tilde{A}+\tilde{B})\frac{t_3}{t_4} & = & 4\,  n_2 \\ \vspace*{.2cm}
(\tilde{A}-\tilde{B})\frac{1}{t_4} & = & 4\,  n_3\\ \vspace*{.2cm}
(\tilde{A}-\tilde{B})\frac{r_3}{t_4} & = & 4\,  n_4\\ \vspace*{.2cm}
(\tilde{A}-\tilde{B})\frac{r_4}{t_4} & = & 4\,  n_5\\ \vspace*{.2cm}
(\tilde{A}-\tilde{B})\frac{(r_3r_4 +t_3t_4)}{t_4} & = & 4\,  n_6\\ \vspace*{.2cm}
(\tilde{A}+\tilde{B})\frac{(t_3r_4 - r_3t_4)}{t_4} & = & 4\,  n_7
\end{array}
\label{cond2}
\eeq
where $\tau_3 = r_3 + it_3$, $\tau_4=r_4+it_4$, and $n_i \in \inte$. These equations in turn imply
\beq
\begin{array}{rcl}\vspace*{.25cm}
\ta & = & 2 (n_1 + n_3 t_4) \\
\tb & = & 2 (n_1 - n_3 t_4)
\end{array}
\label{conda}
\eeq
\bea \nonumber
\frac{t_3}{t_4} & = & \frac{n_2}{n_1}\ \in\ {\bf Q}\\ \nonumber
r_3 & = & \frac{n_4}{n_3}\ \in\ {\bf Q}\\ \nonumber
r_4 & = & \frac{n_5}{n_3}\ \in\ {\bf Q}\\
t_3 t_4 & = & \frac{n_6}{n_4} - \frac{n_4n_5}{n_3^2}\ \in\ {\bf Q}
\label{condb}
\eea 
and
\beq
n_7 n_3\, =\, n_2 n_5 - n_1 n_4 
\label{condc}
\eeq

From eq.(\ref{conda}) and by using $t_4 = 1/g_s$ we obtain 
\beq
\lambda \equiv \frac{\tb}{\ta} = {1 - g_s n_1/n_3 \over 1 + g_s n_1/n_3}
\label{famousap}
\eeq
which is eq.(\ref{famous}) in the main text. On the other hand, eq.(\ref{condb}) show that the closed string moduli $\tau_3$, $\tau_4$ are stabilized and can only take discrete values which depend on the lattice of flux quanta $\{n_i\}_{i=1}^7$. Finally, eq.(\ref{condc}) shows that this lattice is six dimensional, matching the number of free parameters in our problem.

\section{Appendix II: Fluxes and NSNS tadpoles}

Given a string theory compactification, there is a series of local and global constraints that must be fulfilled in order to yield a consistent theory. In particular, the absence of divergences in the open string sector imposes cancellation of Ramond-Ramond and Neveu-Schwarz tadpoles. On the one hand, the presence of RR tadpoles is related to gauge anomalies either in the low energy effective theory or in the worldvolume of suitable D-brane probes \cite{local}. On the other hand, NSNS tadpoles signal an instability of the vacuum configuration and usually involve effective potentials for some moduli. Whereas RR tadpole cancellation conditions must always be imposed \cite{cai}, NSNS tadpoles may remain uncancelled, the divergences associated with them being cured by a Fischler-Susskind mechanism \cite{FS}. Some supergravity solutions in the presence of these global tadpoles have been analyzed in the literature \cite{dmourad,bfont,dilaton} and, in general, they involve a redefinition of the background where Poincar\'e invariance is spontaneously broken.\footnote{See also \cite{Dudas:2004nd} for recent progress in understanding NSNS tadpoles.} In the limit where the supergravity approximation is valid, RR and NSNS tadpole conditions can be recasted as imposing the Bianchi identities/equations of motion for the supergravity fields. This fact was used in \cite{gkp} in order to show the existence of warped compactifications arising as the low energy supergravity limit of certain type IIB orientifolds.

It turns out that global NSNS tadpoles are indeed present in most of the semi-realistic chiral flux compactification constructed previously \cite{blum,cu,ciu,throat}. In addition, they are a necessary ingredient in any model involving flux-induced SUSY breaking on D3-branes. Finally, they would also be present in the recent proposals for obtaining de Sitter vacua from string theory \cite{kklt,silver}. Of course, these tadpoles are no longer present as such once all the moduli of the compactification have been stabilized

In order to have an estimate of the `amount' of NSNS tadpoles in a given type IIB string compactification, let us then consider type IIB supergravity with action
\beqa \nonumber
S_{IIB} & = & { 1 \over 2\k_{10}^2} \int d^{10}x \sqrt{-g} \left\{\car - {\p_M\tau \p^M\bar{\tau} \over 2(\pim \tau)^2} - {G_3 \cdot \overline{G}_3 \over 12 \pim \tau} - {\tilde{F}_5^2 \over 4\cdot 5!} \right\}
 + { 1 \over 8i \k_{10}^2} \int {C_4 \wedge G_3 \wedge \overline{G}_3 \over \pim \tau} \\
& & + \sum_i \left\{ -\int_{\cS_{p_i}} d^{p+1}\xi\ T_{p_i}\  \sqrt{-g} + \mu_{p_i} \int_{\cS_{p_i}} C_{p+1} \right\}
\label{typeIIB}
\eeqa
where $\tau = a + i/g_s$ is the usual type IIB axion-dilaton coupling, $G_{3} = F_3 - \tau H_3$ stands for the complexified RR $+$ NSNS 3-form flux, $\tilde{F}_5$ is the self-dual five-form field strength and $C_{p+1}$ are the RR $(p+1)$-form potentials. On the second line we have included the contribution given by localized sources, such as D$p$-branes with charge and tension $\mu_{p_i}$ and $T_{p_i}$, respectively, and wrapping a $(p+1)$-submanifold $\cS_{p_i}$ of the $D=10$ target space. 

Now, again in the limit where the supergravity approximation is valid, the amount of NSNS tadpole can be estimated by performing a dimensional reduction of the action (\ref{typeIIB}). That is, we aim to compute the excess of tension coming from a compactification of $D=10$ type IIB supergravity $+$ localized sources. This computation has been performed in \cite{giddings} for the particular case of BPS-like warped compactifications of \cite{gkp}. Let us now consider the more general situation where no specific BPS-like supergravity solution is imposed.\footnote{See \cite{alwis} for related issues in flux compactification.}

We will again consider a metric and five-form ansatze similar to the ones used in \cite{gkp}. Namely,
\beq
ds^2_{10} = e^{2A(y)} g^{(4)}_{\mu\nu} dx^\mu dx^\nu + e^{-2A(y)}\tilde{g}_{mn}dy^m dy^n
\label{metric}
\eeq
which is of the usual warped form, with the warp factor $A$ depending on the internal coordinates $y^m$ of the compact six-manifold $\cam_6$, whereas being independent of the four-dimensional coordinates $x^\mu$. Similarly,
\beq
\tilde{F}_{5} = (1 + *) [ d\alpha \wedge d{\rm Vol}^{(4)} ] 
\label{fiveform}
\eeq
where $\alpha$ is an arbitrary function of the internal coordinates $y^m$, and $d{\rm Vol}^{(4)}$ is the volume form of the four-dimensional unwarped metric $g^{(4)}_{\mu\nu}$.  We also consider D$p$-branes filling the $D=4$ non-compact dimensions, that is, $\cS_{p_i} = \cam_4 \times \Sig_i$, where $\Sig_i$ is a $(p-3)$-cycle of $\cam_6$.  Finally, the field-strength 3-form fluxes $G_3$ and $\overline{G}_3$ will be given by closed 3-forms on $\cam_6$, and the complex dilaton will only depend on internal coordinates $\tau = \tau(y)$.

The contribution coming from the Einstein-Hilbert and five-form terms is given by
\beqa \nonumber
& & { 1 \over 2 \k_{10}^2} \int d^{10}x \sqrt{-g} \left\{ \car - {\tilde{F}_5^2 \over 4 \cdot 5} \right\} \\ 
& & \ \longrightarrow \ { 1 \over 2 \k_{4}^2} \int d^{4}x \sqrt{-g} \left\{ \car^{(4)} - \oh V_w^{-1} \int d^6y \sqrt{\tilde{g}} e^{-10A} \left[ (\p e^{4A})^2 + (\p \a)^2  \right]  \right\}
\label{EH+5}
\eeqa
where we are contracting indices with the warped metric. Here, as in \cite{giddings}, we have defined the warped volume $V_w$ and four-dimensional Newton constant as
\beq
\frac{1}{2\k_4^2} = \frac{V_w}{2\k_{10}^2}, \quad \quad V_w \equiv \int d^6y \sqrt{\tilde{g}} e^{-4A}
\label{volume}
\eeq

In order to simplify our discussion let us consider a compactification where only sources of D3-brane and tension are present. By considering (\ref{EH+5}) as well as the terms for the $G_3$ flux and the localized sources, we can rewrite the type IIB action (\ref{typeIIB}) as\footnote{If we were considering more general compactifications, i.e., involving D7-branes, there would be extra contributions to the action, as the one coming from the dilaton variation on the internal dimensions.}
\beqa 
S_{IIB}^{\rm eff} & = & { 1 \over 2 \k_{4}^2} \int d^{4}x \sqrt{-g} \left\{ \car^{(4)} - V_{\rm eff}\right\} 
\label{reduced} \\ \label{potential}
V_{\rm eff} & = & V_w ^{-1} \int d^6y \sqrt{\tilde{g}} \left[\oh e^{-8A} \left[ (\p e^{4A})^2 + (\p \a)^2  \right] +  e^{-2A} { G_3 \cdot \overline{G}_3 \over 12 \pim \tau} \right. 
\\ \nonumber
& & \hspace*{2.75cm} \left.
+ 2 \k_{10}^2 T_3 e^{-2A} \sum_i t_i \delta(y_i) \right] 
\\ \label{potential2}
& = & V_w ^{-1} \int d^6y \sqrt{\tilde{g}}  \left[\oh e^{-10A} \left[\p (e^{4A} \pm \a)\right]^2 + e^{-2A} {\left| i G_3 \pm *_6 G_3\right|^2 \over 24 \pim \tau} \right. 
\\ \nonumber
& & \hspace*{2.75cm} \left. + 2 \k_{10}^2 T_3 e^{-2A} \sum_i \left[t_i \delta(y_i) \pm \left( \rho_{3}^{\rm loc} \right)_i \right] \right]
\eeqa
where $y_i$ are the positions of the D3-branes/O3-planes in $\cam_6$, and $t_i$ are the units of tension of each object, normalized such that a D3-brane has $t = 1$ in the covering space.\footnote{In order to relate (\ref{potential}) and (\ref{potential2}) we have made use of the result
\beq
e^{-10A} \p_m e^{4A} \p^m \a \sim - i e^{-2A} { G_3 \cdot *_{6} \overline{G}_3 \over 12 \pim \tau} - 2 \k_{10}^2 e^{-2A} T_3 \sum_i \left( \rho_{3}^{\rm loc} \right)_i
\label{bianchi2}
\eeq
deduced from the Bianchi identity of $F_5$. Here $\sim$ stands for equality up to integration on $\cam_6$.}

It easy to see that $V_{\rm eff}$ identically vanishes if we impose the conditions
\beqa\label{BPS1}
T_3 t_i & = & \eps\ T_3 \left( \rho_{3}^{\rm loc} \right)_i \\ \label{BPS2}
\ *_6 G_3 & = & \eps\ i G_3 \\
e^{4A} & = & \eps\ \a
\label{BPS3}
\eeqa
for a definite choice of sign $\eps = \pm$. Eqs.~(\ref{BPS1}-\ref{BPS3}) with the choice $\eps=+$ are (part of) the supergravity BPS-like solution found in \cite{gkp}. The choice $\eps = -$ corresponds to nothing but a BPS-like solution with the opposite choice of supersymmetry inside the $\cn = 2$ parent theory.

As explained in \cite{gkp}, conditions (\ref{BPS1}) comes from the saturation of a BPS-like bound on the localized sources. Eq.(\ref{BPS1}) with $\eps = +$ is satisfied by objects with same charge and tension, such as D3-branes and usual O3-planes, whereas objects with opposite charge and tension, such as anti-D3-branes, satisfy it for $\eps = -$. Similarly, (\ref{BPS2}) can be seen as a BPS-like condition for the 3-form flux. More precisely, the ISD condition ($*_6 G_3 =  i G_3$) corresponds to $G_3$ carrying the charge and tension of $|Q_{\rm flux}|$ D3-branes, while a IASD flux ($*_6 G_3 = - i G_3$) carries the RR and NSNS charges of $|Q_{\rm flux}|$ anti-D3-branes. Notice that given a well-quantized 3-form flux $G_3$, it may attain either the ISD or the IASD condition, but never both. 

In general, if we consider flux compactifications with BPS-like localized objects and fluxes, satisfying (\ref{BPS1}) and (\ref{BPS2}), but with different choice of $\eps$ for any of them, we will get a compactification with non-vanishing, positive $V_{\rm eff}$, which indicates the presence of a NSNS tadpole. In the following we will take the value of $V_{\rm eff}$ as a measure of this excess of tension.

Let us consider the particular case where only sources of D3-brane and tension are present. The expression for $V_{\rm eff}$ simplifies to:
\beqa \label{potD3}
%\nonumber
V_{\rm eff} & = & V_w ^{-1} \int d^6y \sqrt{\tilde{g}}  \left[\oh e^{-8A} \left[ (\tilde{\nabla} e^{4A})^2 + (\tilde{\nabla} \a)^2  \right] +  e^{4A} { G_3 \cdot \overline{G}_3 \over 12 \pim \tau}
+ 2 \k_{10}^2 T_3 e^{4A} \sum_i t_i \tilde{\d}(y_i) \right] \\ \nonumber
 & = & V_w ^{-1} \int d^6y \sqrt{\tilde{g}}  \left[\oh e^{-8A} \left[\tilde{\nabla} (e^{4A} \pm \a)\right]^2 + e^{4A} {\left| i G_3 \pm *_6 G_3\right|^2 \over 24 \pim \tau} + 2 \k_{10}^2 T_3 e^{4A} \sum_i  (t_i \pm q_i) \tilde{\d}(y_i) \right]
\eeqa
where $y_i$ are the positions of the D3-branes and O3-planes in $\cam_6$, and $t_i$, $q_i$ are the units of tension and charge of these objects, normalized such that a D3-brane has $t = q = 1$ in the covering space of the orientifold. Here the delta function $\tilde{\d}$ is defined in terms of the unwarped metric $\tilde{g}_{mn}$, and the same for the scalar products, so that we are extracting all the warp factor dependence.

In order to illustrate the use of this formula, let us first consider a system of a D3 and anti-D3-brane, located at $y_{D3}$ and $y_{\overline{D3}}$. Eq.(\ref{potD3}) reads
\beqa
V_{\rm eff} & = & V_w ^{-1} \int d^6y \sqrt{\tilde{g}}  \oh e^{-8A} \left[ (\tilde{\nabla} e^{4A})^2 + (\tilde{\nabla} \a)^2  \right] 
+ T_3 \left( e^{4A(y_{D3})} + e^{4A(y_{\overline{D3}})} \right)
\label{D3antiD3} \\
& = & V_w ^{-1} \int d^6y \sqrt{\tilde{g}} \oh e^{-8A} \left[\tilde{\nabla} (e^{4A} - \a)\right]^2 + 2 T_3 e^{4A(y_{\overline{D3}})} \\
& = & V_w ^{-1} \int d^6y \sqrt{\tilde{g}} \oh e^{-8A} \left[\tilde{\nabla} (e^{4A} + \a)\right]^2 + 2 T_3 e^{4A(y_{D3})}.
\eeqa
Notice that the first contribution in these expressions is always non-vanishing, since the relation (\ref{BPS3}) between the 5-form and the warp factors only holds in the BPS-like compactifications of \cite{gkp}. On the other hand, its contribution is suppressed by the (warped) volume of the compact manifold. The second contribution is also suppressed by the warp factor $e^{4A}$, which is usually minimized in the location of objects of positive tension. 

In the case of an anti-D3-brane in a ISD 3-form flux background $G_3$ carrying the charge of a single D3-brane, we have instead.
\beqa
V_{\rm eff} & = & V_w ^{-1} \int d^6y \sqrt{\tilde{g}}  \oh e^{-8A} \left[ (\tilde{\nabla} e^{4A})^2 + (\tilde{\nabla} \a)^2  \right] 
+ T_3 \left(N_{\rm flux}^w + e^{4A(y_{\overline{D3}})} \right)
\label{fluxantiD3} \\
& = & V_w ^{-1} \int d^6y \sqrt{\tilde{g}} \oh e^{-8A} \left[\tilde{\nabla} (e^{4A} - \a)\right]^2 + 2 T_3 e^{4A(y_{\overline{D3}})} \\
& = & V_w ^{-1} \int d^6y \sqrt{\tilde{g}} \oh e^{-8A} \left[\tilde{\nabla} (e^{4A} + \a)\right]^2 + 2 T_3 N_{\rm flux}^w.
\eeqa
where
\beq
N_{\rm flux}^w  =  - {1 \over (4\pi^2 \a^\prime)^2} \int_{\cam_6} e^{4A} {G_3 \wedge *_6 \overline{G}_3 \over 2\pim \tau} 
\label{warpedtensionflux}
\eeq
A similar expression applies to a set of D3-branes in presence of a IASD $G_3$ flux. We hence see that  D3-branes $+$ IASD fluxes and anti-D3-branes $+$ ISD fluxes, may be on equal footing from the point of view of NSNS tadpoles, thus Einstein's equations of motion. At least once the backreaction of the localized objects has been taken into account. In both cases, the contribution to the excess of tension strongly depends on the warp factor of the compactification.

\end{document}